\documentclass[11pt,a4paper]{article}
\usepackage{jheppub}
\usepackage{verbatim}
\input epsf.tex

\def\slashchar#1{\setbox0=\hbox{$#1$}           
   \dimen0=\wd0                                 
   \setbox1=\hbox{/} \dimen1=\wd1               
   \ifdim\dimen0>\dimen1                        
      \rlap{\hbox to \dimen0{\hfil/\hfil}}      
      #1                                        
   \else                                        
      \rlap{\hbox to \dimen1{\hfil$#1$\hfil}}   
      /                                         
   \fi}

    \def\slashword#1{\setbox0=\hbox{$#1$}        
  \dimen0=\wd0                                   
   \setbox1=\hbox{/} \dimen1=\wd1                
   \ifdim\dimen0>\dimen1                         
      \rlap{\hbox to \dimen0{\hfil\bf---\hfil}} %
      #1                                         %
   \else                                         
      \rlap{\hbox to \dimen1{\hfil$#1$\hfil}}    
      /                                          
    \fi}                                         %

\catcode`@=11
\newdimen\vbigd@men

\def\vbig#1#2{{\vbigd@men=#2\divide\vbigd@men by 2%
   \hbox{$\left#1\vbox to \vbigd@men{}\right.\n@space$}}}
\catcode`@=12

\catcode`@=11
\def\citenum#1{\csname b@#1\endcsname}
\catcode`@=12

\def\lsim{\raisebox{-.1em}{$
\buildrel{\scriptscriptstyle <}\over{\scriptscriptstyle\sim}$}}
\def\gsim{\raisebox{-.1em}{$
\buildrel{\scriptscriptstyle >}\over{\scriptscriptstyle\sim}$}}

\def\dofig#1#2{\centerline{\epsfxsize=#1\epsfbox{#2}}}

\title{Improving SUSY Spectrum Determinations at the LHC
with the Wedgebox Technique}

\author[a]{Ran Lu}
\author[b]{Mike Bisset}
\author[c]{Nick Kersting}

\affiliation[a]{Department of Physics, University of Michigan}
\affiliation[b]{Department of Physics, Tsinghua University, P.R. China}
\affiliation[c]{Physics Department, Sichuan University, P.R. China}

\emailAdd{luran@umich.edu}
\emailAdd{bisset@mail.tsinghua.edu.cn}
\emailAdd{nkersting@scu.edu.cn}

\abstract{
The LHC has the potential not only to discover supersymmetry (SUSY),
but also to permit fairly precise measurements of at least a portion of
the sparticle spectrum.
Proposed mass reconstruction methods rely upon either inverting
invariant mass endpoint expressions or upon solving systems of mass-shell
equations.
These methodologies suffer from the weakness that one certain specific
sparticle decay chain is assumed to account for all the events in the sample.
Taking two examples of techniques utilizing
mass-shell equations, it is found that also applying the wedgebox technique 
allows for the isolation of a purer event sample,
thus avoiding errors, possibly catastrophic, due to mistaken assumptions
about the decay chains involved and simultaneously improving accuracy.
What is innovative is using endpoint measurements
(via the wedgebox technique)
to obtain a more homogeneous, well-understood sample set
rather than just using said endpoints to constrain the values of the masses
(here found by the mass-shell technique).
The fusion of different established techniques in this manner represents a
highly profitable option for LHC experimentalists who will soon have data to
analyze.
}


\arxivnumber{0806.2492}

\begin{document}

\maketitle


\section{Introduction}

LHC experimentalists will soon determine whether or not
Supersymmetry (SUSY) is a TeV-scale phenomenon: if so, colored sparticles
will probably be the first to be discerned, possibly soon followed by
neutralinos, charginos, and sleptons if favorable decay channels are open,
though measuring the masses of these latter colorless sparticles with percent
precision will be challenging\cite{:1999fr,Ball:2007zza}. The reason for this is that
every R-parity-conserving\footnote{If R-parity is not conserved, then it may
be possible to fully reconstruct events.  See \cite{Allanach:2001xz} for further
details.} SUSY event produces at least two invisible particles
(the lightest SUSY particles, or LSPs) which carry away significant missing
energy and make it impossible to reconstruct mass peaks. Therefore,
many SUSY mass extraction techniques depend on precise measurement of
invariant mass distribution endpoints.  For a sparticle decaying into an
LSP and a Standard Model (SM) fermion pair, either via a three-body decay or
sequential two-body decays, it is straight-forward to see how the endpoint of
the di-fermion invariant mass distribution yields the mass difference between
the decaying sparticle and the LSP (perhaps modified by the on-mass-shell
intermediate for two-body decays) \cite{Paige:1996nx,Hinchliffe:1996iu,Gianotti:682494,Hinchliffe:1998ys,Hinchliffe:1999zc,Mura:1311179}.  Studies
attempting to fully reconstruct the actual sparticle masses from invariant
mass endpoint information rely on {\em specific} longer decay chains, typically
$\widetilde{q} \to \widetilde{\chi}^0_2 q \to
\widetilde{\ell}^{\pm} \ell^{\mp} q \to
\widetilde{\chi}^0_1 \ell^+ \ell^- q$
\cite{Bachacou:1999zb,Gjelsten:2004ki,Gjelsten:2005aw,Gjelsten:2005sv,Gjelsten:2005vv,Gjelsten:708246,Gjelsten:2006as,Miller:2005zp,Lytken:703773,Butterworth:2007ke,Lester:2006cf,Allanach:2000kt,Tovey:2003ef,Borjanovic:2005tv,Tovey:2008ui} ---
each event would have two $\tilde{q}$'s (for instance) produced.
It is then theoretically doable to construct enough invariant mass
distributions to determine the sparticle masses; however, in practice
endpoint measurement may be complicated by low event rates, fitting criteria,
unaccounted-for (in the simulation) higher-order and radiative\cite{Horsky:2008yi}
effects, and (in particular) backgrounds.

Ideas on how to measure SUSY masses without relying on distribution endpoints
have also been put forward.
The work of Nojiri {\it et al.}\cite{Nojiri:2003tu,Kawagoe:2004rz}, for example,
uses mass-shell relations in a sufficiently long SUSY decay chain, {\it e.g.}
$\tilde{g} \to b ~\tilde{b}  \to \widetilde{\chi}_{2}^0 bb
                             \to \widetilde{l}^\pm l^\mp bb
                         \to l^\pm l^\mp bb  \widetilde{\chi}_{1}^0$;
if some of the masses are already known in this chain the others
can  \emph{in principle} be found by solving mass-shell relations for a
small sample of events\footnote{In practice, many events are still required.},
which may in fact lie far from the endpoint.
Another method, due to Cheng {\it et al.}\cite{Cheng:2007xv,Cheng:2008mg,Cheng:2009fw}, starts from very
similar looking mass-shell relations, but instead of assuming some masses and
solving for the others, scans the whole mass-space for points where these
relations are most likely to be satisfied. Both methods, hereafter
designated as `Mass Shell Techniques' (MSTs)\footnote{Refs.
\cite{Nojiri:2003tu,Kawagoe:2004rz,Nojiri:2007pq} use the name
`mass relation method' to refer to their technique.}, seem quite effective
in obtaining percent-level determination of the sought-after SUSY masses,
at least at the parameter points considered in those works.

The accuracy of both these MSTs hinges on one critical assumption:
the decay topology of choice has been isolated.
In the actual LHC data, the decay topology would have to be inferred, if this
is at all possible, before proceeding; MSTs would thus appear to be
excellent roads to SUSY mass reconstruction which, however, begin only at a
point half-way to the destination.

The present work focuses on the first half of this road; {\it i.e.},
isolating a desired decay topology\footnote{The recent paper \cite{Bai:2010hd} addresses similar issues from a somewhat different perspective.}, and on how this affects a subsequent MST analysis.  As a first foray into this potentially quite thorny task, consider specific topologies studied in \cite{Nojiri:2003tu} and \cite{Cheng:2007xv,Cheng:2008mg,Cheng:2009fw} involving a pair of neutralinos
$\widetilde{\chi}_{i}^0 \widetilde{\chi}_{j}^0$ ($i,j=2,3,4$) that subsequently decay to leptons (electrons and muons) via on-shell sleptons.
This situation is amenable to a wedgebox analysis \cite{Bisset:2005rn,Bian:2006xx}
which is based upon a scatter plot of the di-electron mass $M_{e^+ e^-}$
versus the dimuon mass $M_{\mu^+ \mu^-}$.
A key benefit of this technique is that it allows (at least partial)
separation of individual events according to the specific $(i,j)$-pair
whose production gave rise to them.  Given sufficient statistics, events
from each such decay-type fall in distinct, easily-recognized zones of the
wedgebox plot.
The overall topology of the resulting wedgebox plot then tests for the
significant presence of the various possible $(i,j)$ decay channels --- which
may for instance signal the meaningful presence of a decay channel
erroneously assumed to have been insignificant as the basis for a MST
analysis.  Events can be selected from a specific zone of the wedgebox plot,
preferably a zone populated by only one decay channel.  This acts to
maximize sample homogeneity and assure the basic MST assumption is satisfied.

Although the wedgebox technique relies on locating the endpoints of
invariant mass distributions --- just like the studies
\cite{Bachacou:1999zb,Gjelsten:2004ki,Gjelsten:2005aw,Gjelsten:2005sv,Gjelsten:2005vv,Gjelsten:708246,Gjelsten:2006as,Miller:2005zp,Lytken:703773,Butterworth:2007ke,Lester:2006cf,Allanach:2000kt,Tovey:2003ef,Borjanovic:2005tv,Tovey:2008ui} mentioned
earlier, the information sought is radically different:  the wedgebox analysis
is tailor-made for event sample sets comprised of assorted produced sparticle
pairs and multiple sparticle decay chains.
The observed endpoints serve to delineate the zones and allow for
selection of purer subset(s) from an overall sample set.
(Using this endpoint information to determine a set of cuts is
a far more rational course than that of arbitrarily choosing some
numerical cut-off values to purify the data sample.)
Virtually all previous studies \emph{presume} such purification has already
been accomplished either by an unspecified set of cuts or a fortuitous
choice of SUSY input parameters.  The wedgebox technique illustrates a
concrete method of how to deal with the more general case, and
the consequences which result and should not be ignored in a coupled
analysis aiming to extract the sparticle masses.

This paper will show that MSTs, by construction unrelated to invariant
mass endpoints, can nevertheless be improved by information contained in
these endpoints --- specifically via the wedgebox technique\footnote{Though 
our strategy is quite different from the `hybrid' method of \cite{Nojiri:2007pq} 
which couples an MST with \emph{values} for of endpoints.}.
The paper is organized as follows: Sections \ref{sec:noj} and
\ref{sec:cheng} concentrate on the `N-MST' method of Nojiri {\it et al.} and
the `C-MST' method of Cheng {\it et al.},
respectively; Section \ref{sec:conc} then offers conclusions
and some additional discussion on SUSY mass spectral analyses at the LHC.

\section{The N-MST Method of Nojiri {\it et al.}}
\label{sec:noj}
\begin{figure}[!thb]
\begin{center}
\dofig{3.15in}{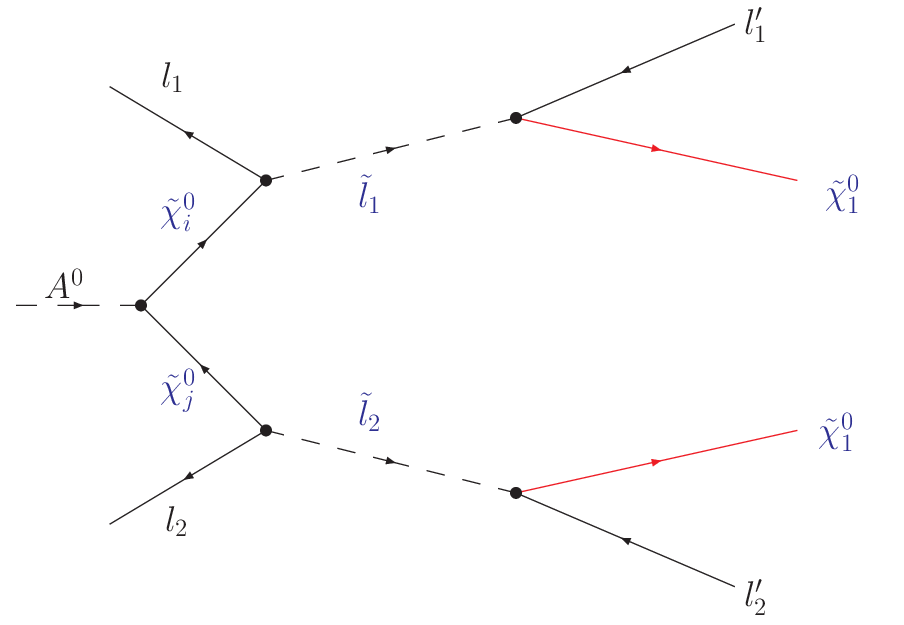}
\end{center}
\vskip -0.6cm
\caption{\small \emph{ Feynman diagram for the decay (\ref{Adecay}).
Here $i,j=2,3,4$ while the label 1 or 2 on leptons stands for either
$e$ or $\mu$.}
 \label{fig:feyn} }
\end{figure}

For the N-MST method, the focus will be on the decay of a
heavy MSSM pseudoscalar Higgs boson as considered in \cite{Nojiri:2003tu}.
The specific decay chain is
\begin{eqnarray}
   pp \to A^0 \to \widetilde{\chi}_{i}^0 \widetilde{\chi}_{j}^0
   \to \widetilde{l}_1^\pm {l_1}^\mp \widetilde{l}_2^\pm {l_2}^\mp
      \to {l_1}^\mp {l'_1}^\pm {l_2}^\mp {l'_2}^\pm
          \widetilde{\chi}_{1}^0 \widetilde{\chi}_{1}^0
\label{Adecay}
\end{eqnarray}
where the Higgs boson decays to neutralinos ($i,j=2,3,4$) via
on-shell sleptons of the electron- or muon- variety (see Fig. \ref{fig:feyn}).
Assuming the final leptons' four-momenta are known while the LSP's escape
detection, (\ref{Adecay}) implies six mass-shell constraints on the
eight unknown components of LSP four-momenta ($p^\mu$ and  $p'^\mu$),
\begin{eqnarray}
\label{con1}
m_{\widetilde{\chi}_1^0}^2 &=& p^2 \\
m_{\widetilde{\chi}_1^0}^2 &=& p'^2 \\
m_{\widetilde{l}_1}^2 &=& (p_{l_1} + p)^2 \\
m_{\widetilde{l}_2}^2 &=& (p_{l_2}+p')^2 \\
m_{\widetilde{\chi}_i^0}^2 &=& (p_{l_1} + p_{l'_1} + p)^2 \\
m_{\widetilde{\chi}_j^0}^2 &=& (p_{l_2} + p_{l'_2} + p')^2
  \label{con1end}
\end{eqnarray}
Nojiri {\it et al.} also posit two overall momentum conservation constraints
\begin{eqnarray}
  (p_{l_1} + p_{l_2} + p_{l'_1} + p_{l'_2} + p + p')\cdot \vec{x}  &=& 0 \\
  (p_{l_1} + p_{l_2} + p_{l'_1} + p_{l'_2} + p + p')\cdot \vec{y}  &=& 0
  \label{con2}
\end{eqnarray}
along directions ($x$ , $y$) transverse to the beam (the $z$-direction), 
though this would appear to be contingent on the Higgs boson having no 
transverse momentum.  If all the masses
$m_{\widetilde{\chi}_1^0}$, $m_{\widetilde{l}_1}$, $m_{\widetilde{l}_2}$,
$m_{\widetilde{\chi}_i^0}$, and $m_{\widetilde{\chi}_j^0}$ are known in
advance, one can solve the eight equations (\ref{con1})-(\ref{con2}) for
the eight unknowns and reconstruct the Higgs boson mass via
\begin{equation}\label{higgscon}
     m_{A}^2 = (p_{l_1} + p_{l_2} + p_{l'_1} + p_{l'_2} + p + p')^2
\end{equation}
from just one Higgs boson event of the type (\ref{Adecay}).
However, even in this idealized scenario which does not include detector
resolution, particle widths, backgrounds, {\it etc.}, there are two major
caveats which prevent this procedure from being so straightforward:
\begin{itemize}
\item There is a 4-fold ambiguity in assigning labels $l_{1,2}, l'_{1,2}$
  to the leptons; this forms a combinatoric background.
\item Combining (\ref{con1})-(\ref{con2}) leads to a quartic equation
  with 0, 2 or 4 solutions for the unknown momenta.
\end{itemize}
So what one must do in practice is collect a number of events and deduce the
correct value of $m_A$ from the maximum of the resulting distribution.
In \cite{Nojiri:2003tu} a $10^3$ event sample (with no backgrounds) thus yielded
a percent-level determination of the Higgs boson mass.

\subsection{Addition of the Wedgebox Technique}
\label{sec:add}
\subsubsection{Box Topology}
As shown in \cite{Nojiri:2003tu} the programme sketched in the previous section
works fairly well at Snowmass Benchmark SPS1a \cite{Allanach:2002nj} where the
dominant Higgs boson decays are via (\ref{Adecay}) with $i=j=2$.
Therein, 1000 events of the type
\begin{equation}\label{NMST-eqn}
     pp \to A^0 \to {\widetilde\chi}^0_2 {\widetilde\chi}^0_2
               \to llll {\widetilde\chi}^0_1 {\widetilde\chi}^0_1
\end{equation}
were generated using the HERWIG6.4
 \cite{Corcella:2000bw,Corcella:2002jc,Moretti:2002eu} event generator
and passed through the detector simulator package ATLFAST \cite{Richter-Was:683751}.
The only cuts required were that all four \emph{isolated}
leptons have $\eta < 2.5$, with two of the leptons also having
$p_T ~> ~ 20\, \hbox{GeV}$
while the other two have $p_T ~> ~ 10\, \hbox{GeV}$.
Same-flavor events such as $e^+ e^- e^+ e^- $ or $\mu^+ \mu^- \mu^+ \mu^- $
were also included if one of the two possible pairings of OS leptons in such
configurations gave a di-lepton invariant mass beyond the
$\sim 78\, \hbox{GeV}$ kinematic endpoint (implying the other possible pairing
is the correct one).  Thirty percent of the original 1000 events passed the cuts and selection criteria, leaving 300 events for the N-MST analysis, and these in turn  
yielded the correct Higgs boson mass with a resolution of only $6\, \hbox{GeV}$ \cite{Nojiri:2003tu}.

The following set of cuts are herein adopted in a effort to reproduce these
results using ISAJET\cite{Baer:1999sp,Paige:2003mg} and a detector
simulation which assumes a typical LHC experiment, as provided by 
private programs checked against results in the literature, in place of HERWIG
and ALTFAST\footnote{The detector simulation of the calorimetry is based on a cell 
size of $\Delta \eta \, \times \Delta \phi = 0.1 \times 0.1$. 
Particle resolutions, adopted to approximate the CMS detector, are given by
$\frac{\Delta p_i}{p_i} \propto
\sqrt{ r_1^2 / p_i + r_2^2}$ where
$r_1 (r_2) = 0.8 (0.03)$ for muons or hadrons with $|\eta| < 2.6$,  
$1.0 (0.05)$ for muons or hadrons with $2.6 \le |\eta| < 4.0$;
and $r_1 (r_2) = 0.03 (0.005)$ for electrons and photons with $|\eta| < 4.0$.
This simulation was also repeated using PYTHIA 6.4 \cite{Sjostrand:2006za} coupled with PGS 4 \cite{PGS4}, yielding results very close to the ISAJET study.
The CTEQ 6.1M \cite{Stump:2003yu} set of parton distribution functions is used with top and bottom quark masses set to $m_t=175\, \hbox{GeV}$ and $m_b=4.25\, \hbox{GeV}$, respectively.}:
\begin{enumerate}
  \item  Leptons must be \emph{hard} ($p_T^\ell > 10, 8\, \hbox{GeV}$
for $e^{\pm},\mu^{\pm}$, respectively; $|\eta^{\ell}|<2.4\,$), and
\emph{isolated}
(no tracks of other charged particles in a $r = 0.3\, \hbox{rad}$ cone
around the lepton, and less than $3\, \hbox{GeV}$ of energy deposited into the
electromagnetic calorimeter for $0.05\, \hbox{rad} < r < 0.3\, \hbox{rad}$
around the lepton).
\item There must be missing energy in the range:
    $20\, \hbox{GeV} < \slashchar{E_{T}} < 130\, \hbox{GeV}\;$.
\item No jets\footnote{Jets are defined by a cone algorithm with $r = 0.4$
and must have $|\eta_j| < 2.4$.} are present with a reconstructed energy
$E_{jet}$ greater than $50\, \hbox{GeV}$.
\newline
\end{enumerate}
A sufficient sample of $pp \to A^0$ events is collected to represent an
integrated LHC luminosity of $300\, \hbox{fb}^{-1}$, the same integrated luminosity as in \cite{Nojiri:2003tu}, though this study only finds about 200 events of the type
(\ref{NMST-eqn}) for this luminosity.  About 11\% of the generated events pass the cuts and selection criteria\footnote{This rises to about 16\% if the jet and missing energy cuts are excluded, and to around 23.5\% if same-flavor events meeting the criteria of \cite{Nojiri:2003tu} were also to be included.  This is comparable to the 30\% noted in \cite{Nojiri:2003tu}.}

\begin{figure}[!thb]
  \begin{center}
  \begin{minipage}[c]{0.5\textwidth} 
    \centering 
        \includegraphics[width=2.7in]{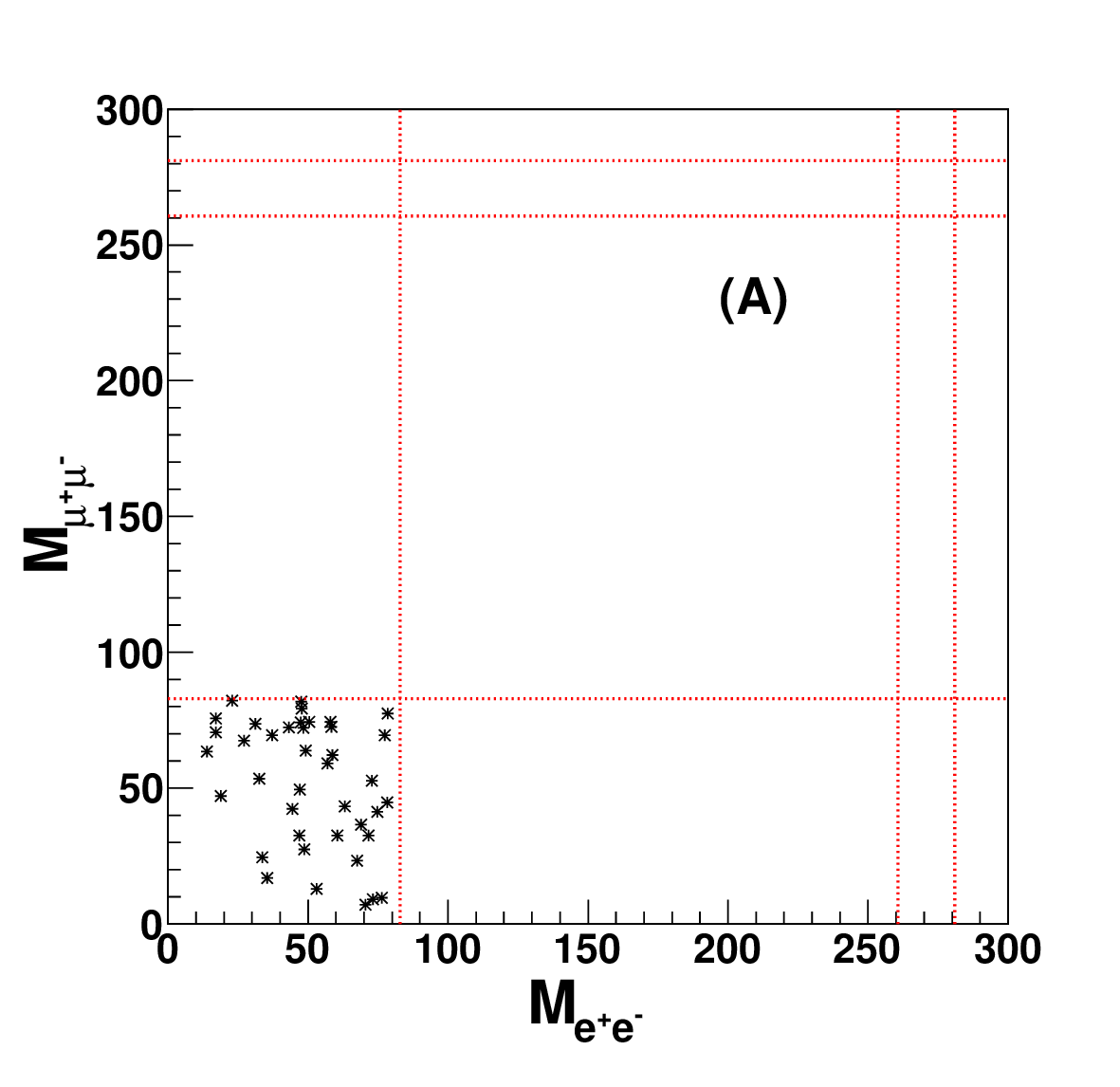}
  \end{minipage}%
  \begin{minipage}[c]{0.5\textwidth} 
    \centering 
        \includegraphics[width=2.7in]{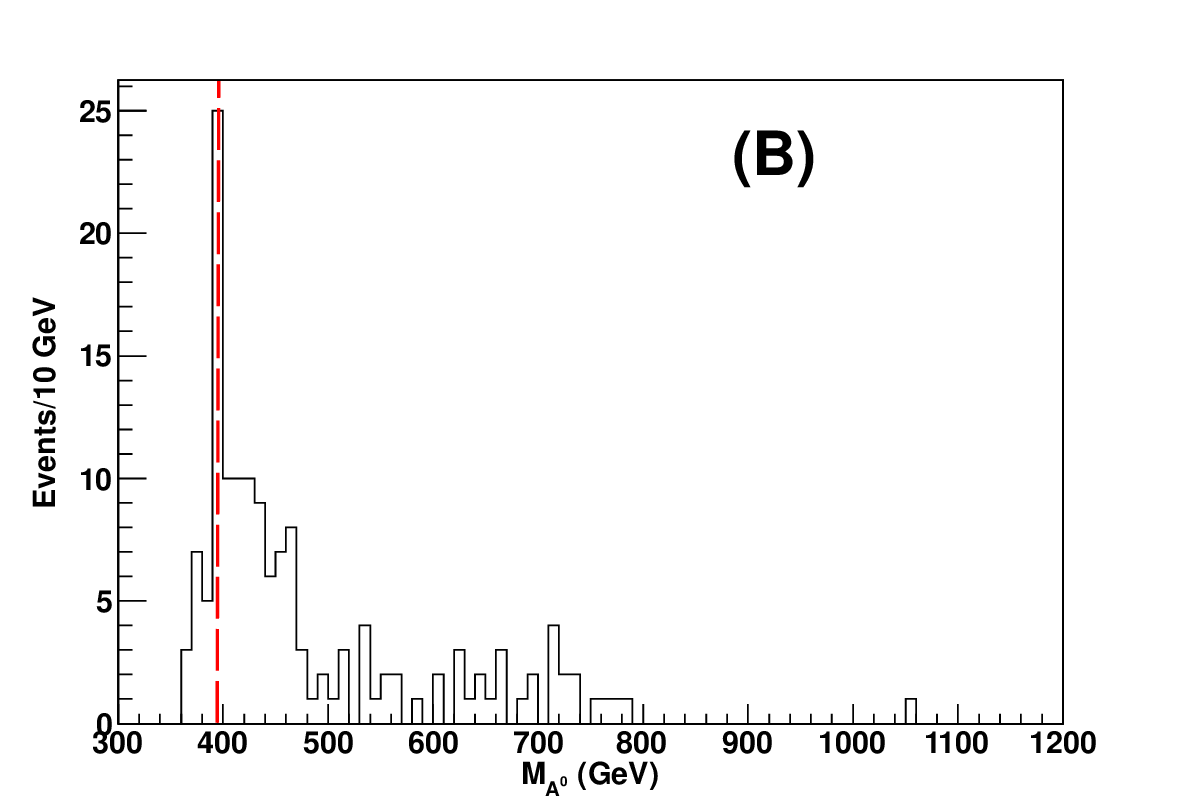}
  \end{minipage}
    \end{center}
\vskip -0.6cm
\caption{ \small \emph{(A) Wedgebox plot at SPS1a ($A^0$ bosons {\em only})
for $300\, \hbox{fb}^{-1}$ luminosity, cuts as in text.  Dotted lines show
locations of kinematic endpoints from ${\widetilde\chi}^0_{2,3,4}
\to {\widetilde\chi}^0_1$ decays. (B) Distribution of solutions for SPS1a
($A^0$ bosons {\em only}).
Here the maximum is at the correct Higgs boson mass of
$m_A \approx 395\, \hbox{GeV}$.}
 \label{fig:sps1a} }
\end{figure}

Fig.~\ref{fig:sps1a}a shows the wedgebox plot at SPS1a
for $p p \to A^0 \to e^+ e^- \mu^+ \mu^-$ events. A `simple box' topology consistent with
the expected ${\widetilde\chi}^0_2 {\widetilde\chi}^0_2$ origin of lepton
pairs is clear.
Moreover, the number of flavor-balanced
events ($e^+ e^- e^+ e^- + \mu^+ \mu^- \mu^+ \mu^- + e^+ e^- \mu^+ \mu^-$)
exceeds the number of flavor-unbalanced
events ($e^+ e^- e^\pm \mu^\mp + \mu^+ \mu^- \mu^\pm e^\mp$)
at this parameter point \cite{Huang:2008qd}; this indicates that the events come
primarily from a Higgs boson decay\footnote{Note that though in a simulation
one can of course choose to only generate Higgs boson decay events,
experimentalists lack this freedom.} with decay topology (\ref{Adecay}).
Though the final number of events passing cuts is somewhat small
(only 40 compared to the $1000 \, \times 0.3 \, = 300$ which \cite{Nojiri:2003tu} estimates, for the reasons noted above) the number of events in the distribution of solutions is quite a bit larger: recall the
bulleted caveats earlier which potentially can yield a
$4 \times 4$ = 16-fold multiplicity factor.
Though only a factor of 4 or so is observed, nonetheless a fairly 
clear peak\footnote{Apparently localized to within the $10\, \hbox{GeV}$
binning size adopted in Fig.~\ref{fig:sps1a}b.} 
in the distribution emerges (see Fig.~\ref{fig:sps1a}b) at the correct
value\footnote{Also note here that the momentum-conservation constraints
(\ref{con2}) do not appear to be generally true; {\it i.e.}, 
according to this analysis the parent Higgs boson is often generated with 
significant transverse momentum in the range 
$0 ~\lsim~ p_T~ \lsim~ 100 \, \hbox{GeV}$.  Surprisingly, this does not seem 
to affect the result.} of $m_A \approx 395\, \hbox{GeV}$.

\begin{figure}[htpb]
    \begin{center}
        \includegraphics[width=2.7in]{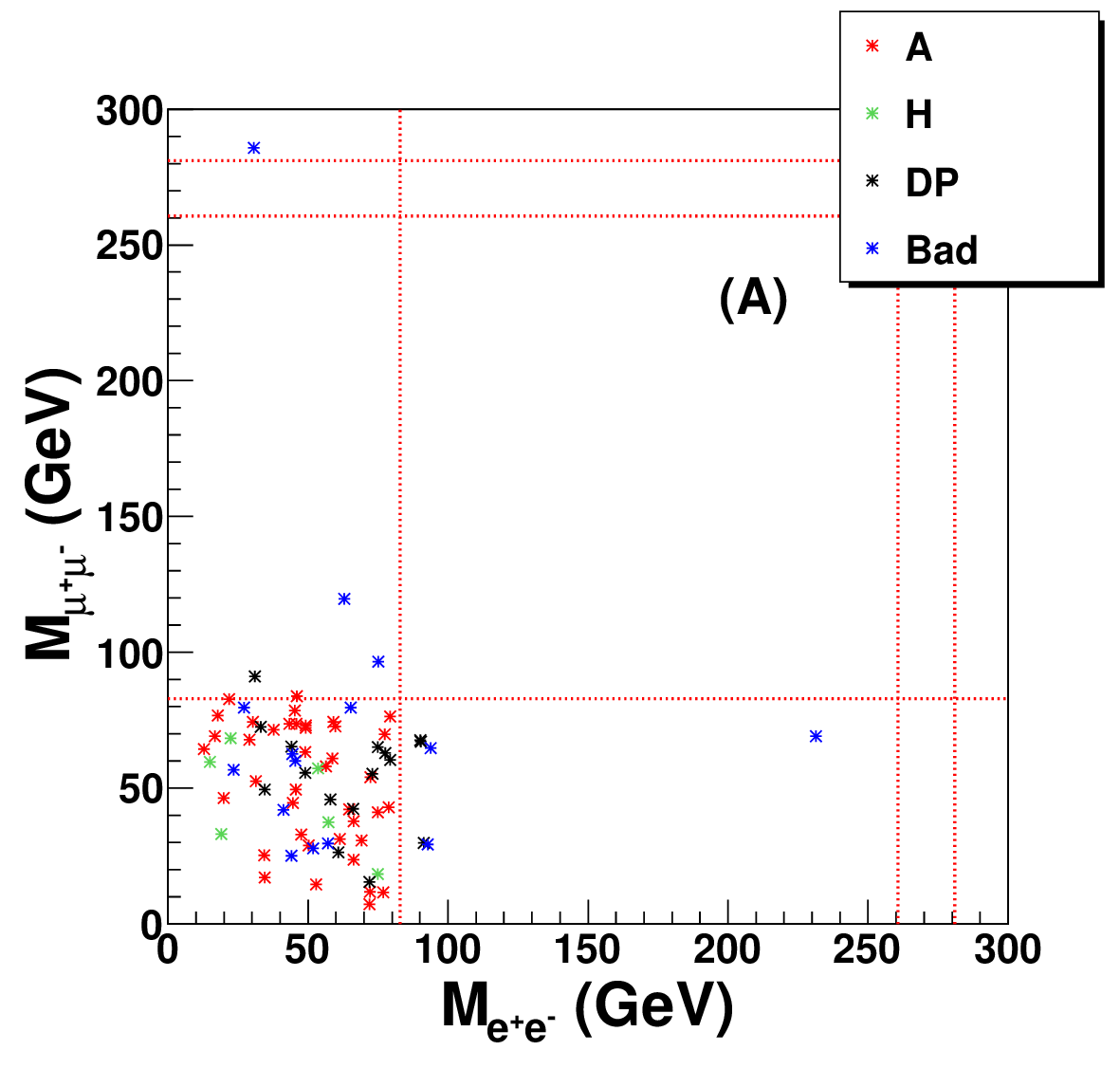}
        \includegraphics[width=2.7in]{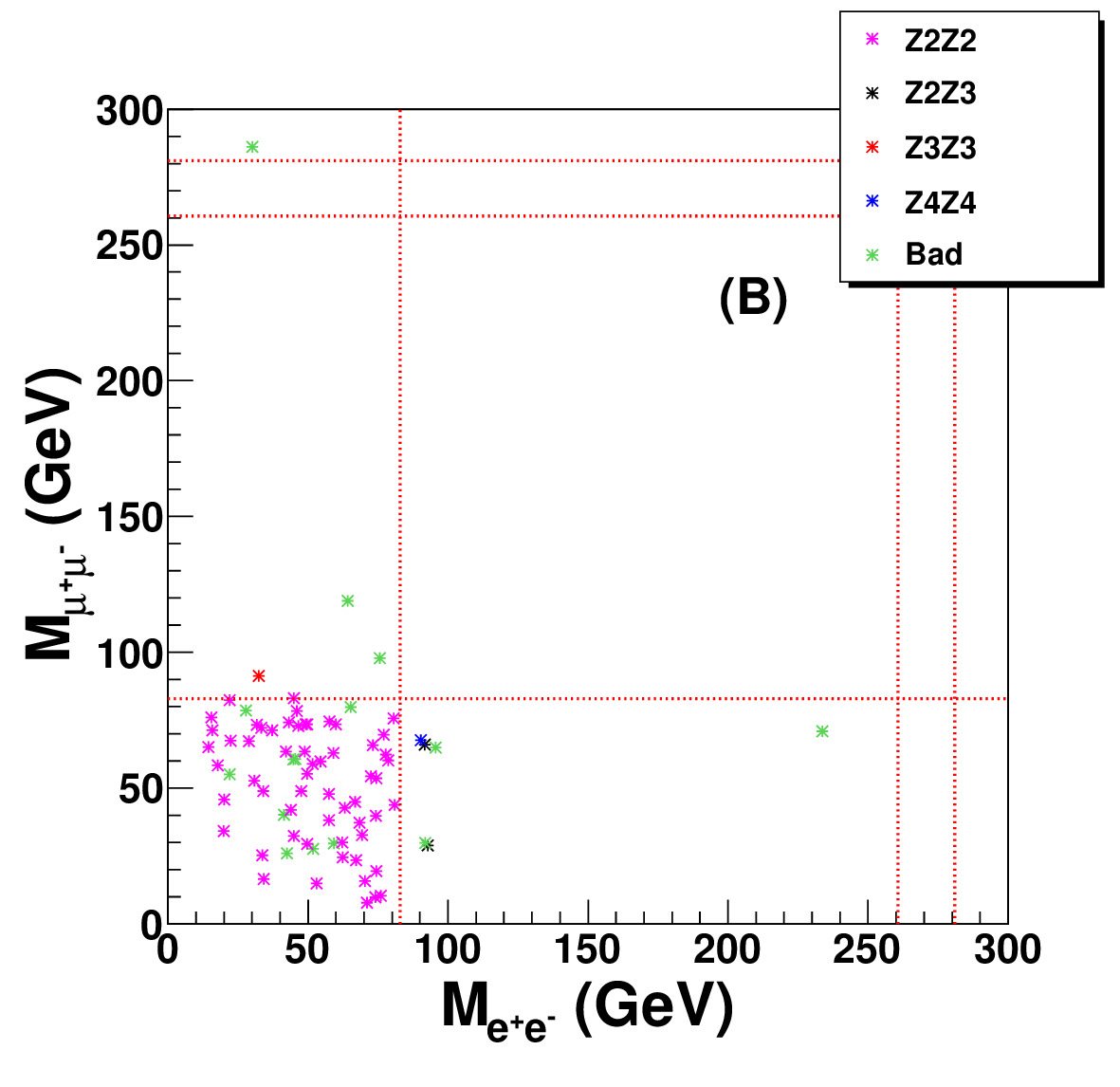}
        \includegraphics[width=2.7in]{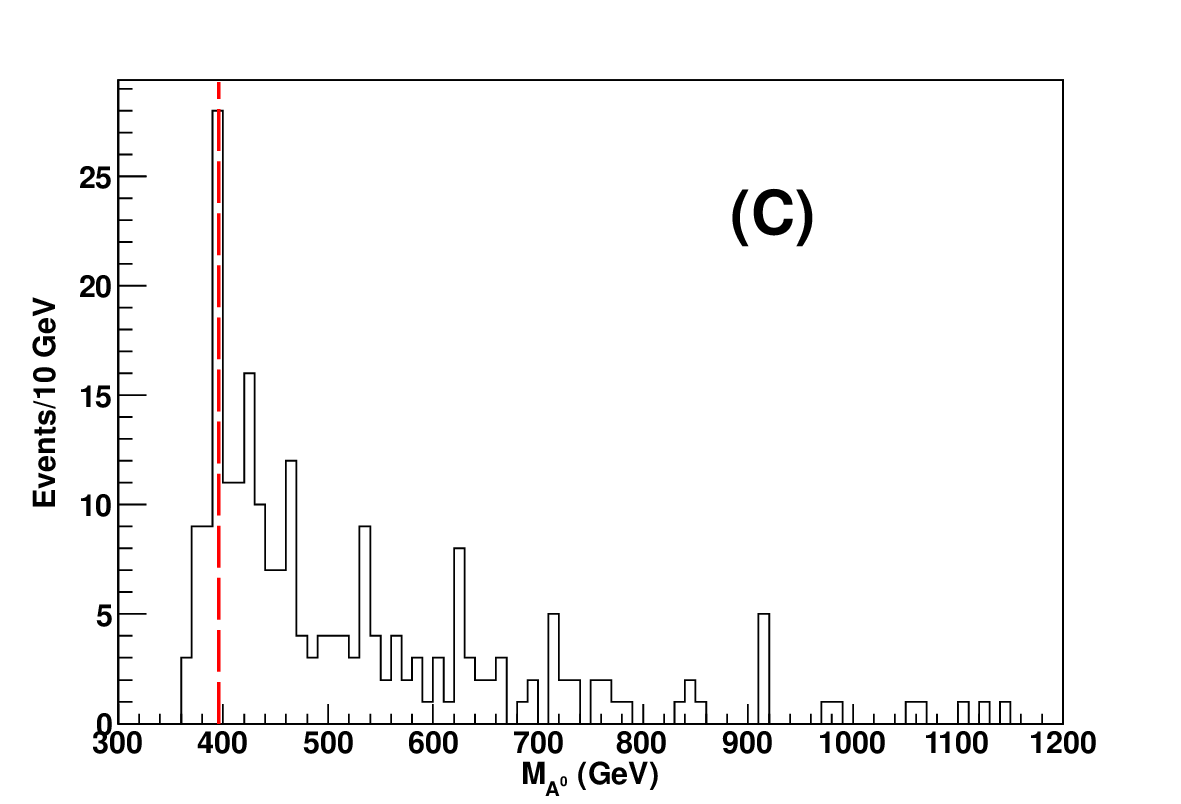}
    \end{center}
    \caption{\small \emph{ (A) Complete wedgebox plot for SPS1a, 
모모for $300\, \hbox{fb}^{-1}$, showing the origin of each 
모모four-lepton event --- labeled as from `$A$', `$H$', `DP'(direct production), or
    otherwise `Bad'.
    (B) specific pair of neutralinos generated in each four-lepton
 event --- identified as  ${\widetilde\chi}^0_2 {\widetilde\chi}^0_2$ (`Z2Z2'),
 ${\widetilde\chi}^0_2 {\widetilde\chi}^0_3$ (`Z2Z3'),
 ${\widetilde\chi}^0_3 {\widetilde\chi}^0_3$  (`Z3Z3'),
 ${\widetilde\chi}^0_4 {\widetilde\chi}^0_4$  (`Z4Z4'),
  or other events-types (`Bad').
(C) Distribution of solutions from full SPS1a event set.
(D) Distribution of solutions from the SPS1a event set 
(inclusion or omission of events lying outside the
${\widetilde\chi}^0_2 {\widetilde\chi}^0_2$-box has virtually 
no impact on this plot).}}
    \label{fig:sps1a-full}
\end{figure}

Naturally, other processes might also generate events of the signal type.  It
is clearly incorrect to consider only events due to $A^0$ production without
also including $H^0$ production.  At SPS1a, $m_A$ and $m_H$ differ by less
than 3 GeV, so the $H^0$ events may be expected to aid in the mass
determination.  However, only 6 $H^0$ events (as compared to 40 $A^0$ events)
are found.  In addition, 16 events from direct neutralino pair production and
15 events from production processes involving charginos (these two event
categories are each produced via an electroweak vector gauge boson) and/or
isolated leptons from heavy-quark decays (in sparticle-containing events) contribute to the background.  Standard Model processes do not contribute any events.
Fig.~\ref{fig:sps1a-full}a shows the wedgebox plot at SPS1a including all
these event-types.  Fig.~\ref{fig:sps1a-full}b then shows the specific
neutralino pair that is generated in each event (irrespective of the
production type).  While the overwhelming majority of the events are due to
${\widetilde\chi}^0_2 {\widetilde\chi}^0_2$ as expected/hoped, a
couple ${\widetilde\chi}^0_2 {\widetilde\chi}^0_3$ events, a
${\widetilde\chi}^0_3 {\widetilde\chi}^0_3$ event, and a 
${\widetilde\chi}^0_4 {\widetilde\chi}^0_4$ event are also present.
Somewhat fortuitously, all four of these 
non-${\widetilde\chi}^0_2 {\widetilde\chi}^0_2$ events fall outside of the 
${\widetilde\chi}^0_2 {\widetilde\chi}^0_2$ box region delineated by the dashed lines in the lower-left corner of Fig.~\ref{fig:sps1a-full}(a,b), as do 6 of the
15 `Bad몶(mainly chargino-containing) events.  Even considering possible additional
experimental smearing, it should be apparent to experimentalists (who do not see
color-coded events) that such events are outside of the ${\widetilde\chi}^0_2 {\widetilde\chi}^0_2$ box.  In this case, the inclusion or removal of these distinguishable wrong-chain events does not markedly alter the distribution of solutions shown in Fig. ~\ref{fig:sps1a-full}c (which is little degraded from the 
$A^0$-only plot shown in Fig.~\ref{fig:sps1a}b).  However, it does indicate how
reference to the exact (within experimental accuracy) numerical value for the edges of the ${\widetilde\chi}^0_2{\widetilde\chi}^0_2$-box delivered by the wedgebox plot (or equivalently by an analysis of the two same-flavor dilepton invariant mass distributions) might aid in improving the N-MST analysis.  The worth of the wedgebox information becomes more pronounced when the topology is not so simple (as is often the case for allowable MSSM parameter sets).  This more general situation is
addressed next.

\subsubsection{Wedge Topology}
The SPS1a parameter point event set is dominated by 
events in which a ${\widetilde\chi}^0_i {\widetilde\chi}^0_j$ $(i=2=j=2)$
pair is generated, consistent with the expected decay chain presumed by the N-MST analysis in \cite{Nojiri:2003tu}.  In principle, this N-MST method should also work for a set of events for which $i \ne j$.  The complication comes when one considers 
that the experimentalists event set (consistent with the cuts already) specified will then in general consist of a mixture of different decay chains involving different
generated pairs of neutralinos.  As a case in point, 
consider the following MSSM parameter set:
\begin{description}
  \item[MSSM Test Point I]
  \begin{eqnarray*}
  \mu = 190\, \hbox{GeV} ~~~ M_2 = 280\, \hbox{GeV} ~~
   \tan \beta = 10  ~~
 M_A = 600\, \hbox{GeV} \\
   M_{{\widetilde{e}, \widetilde{\mu}}_{L,R}} = 150\, \hbox{GeV}~~
   M_{\widetilde{\tau}_{L,R}} = 250\, \hbox{GeV}~~
   M_{\tilde{q},\tilde{g}} = 1000\, \hbox{GeV}.
  \end{eqnarray*}
\end{description}
which has the mass spectrum shown in Table \ref{tab:masses}.
\begin{table}
 \caption{\small \emph{Relevant masses at the MSSM Test Point I,II}}
    \begin{center}
     \begin{tabular}{|c|c|c|} \hline
   Particle & Mass (Point I) (GeV) & Mass(Point II) (Gev) \\ \hline
  ${\widetilde\chi}^0_1$
            & $119.94$ & $86.03$ \\ \hline
  ${\widetilde\chi}^0_2$
            & $180.33$ & $143.09$ \\ \hline
  ${\widetilde\chi}^0_3$
            & $197.98$ & $166.40$\\ \hline
  ${\widetilde\chi}^0_4$
            & $317.72$ & $277.27$ \\ \hline
  $\widetilde{e_R}$
            & $156.17$ & $127.20$ \\ \hline
  $\widetilde{\mu_R}$  & $156.17$ & $127.20$\\ \hline
  $A^0$     & $600.0$  &  $700.00$ \\ \hline
  $H^0$     & $609.28$ &  $705.58$ \\ \hline
       \end{tabular}
    \end{center}
 \label{tab:masses}
\end{table}
Now the main production modes contributing to the $e^+ e^- \mu^+ \mu^- $
signal are the Higgs boson channels
$H^0/A^0 \to {\widetilde\chi}^0_2 {\widetilde\chi}^0_2,
{\widetilde\chi}^0_2 {\widetilde\chi}^0_3,
{\widetilde\chi}^0_2 {\widetilde\chi}^0_4$,
`direct' neutralino pair production channels\footnote{As explained
in \cite{Bian:2006xx}, direct ${\widetilde\chi}^0_2 {\widetilde\chi}^0_2$
channels are suppressed by isospin symmetry, while
${\widetilde\chi}^0_{3,4}{\widetilde\chi}^0_{3,4}$ are
 phase-space suppressed.} ${\widetilde\chi}^0_2 {\widetilde\chi}^0_3
 \; \& \; {\widetilde\chi}^0_2 {\widetilde\chi}^0_4$, and `mixed' channels
involving charginos, mainly
${\widetilde\chi}^\pm_2 {\widetilde\chi}^\mp_2 \; \& \;
{\widetilde\chi}^\pm_2 {\widetilde\chi}^0_{2,3,4}$ production.
A random sample of $e^+ e^- \mu^+ \mu^- $ events\footnote{Note that for
{\bf MSSM Test Point I} the staus have been set more massive than
the other sleptons to avoid tau-containing decays and generate more
decays to the desired leptons.  This is simply done by hand here for
convenience.} will therefore not be a clean collection of Higgs boson decays,
which for the luminosity considered here ($300\, \hbox{fb}^{-1}$) number about 150 against
nearly 800 direct channel and 100 or so mixed channel events.
This point then presents a more challenging case for applying the mass
relation method.

\begin{figure}[htpb]
    \begin{center}
        \includegraphics[width=1.8in]{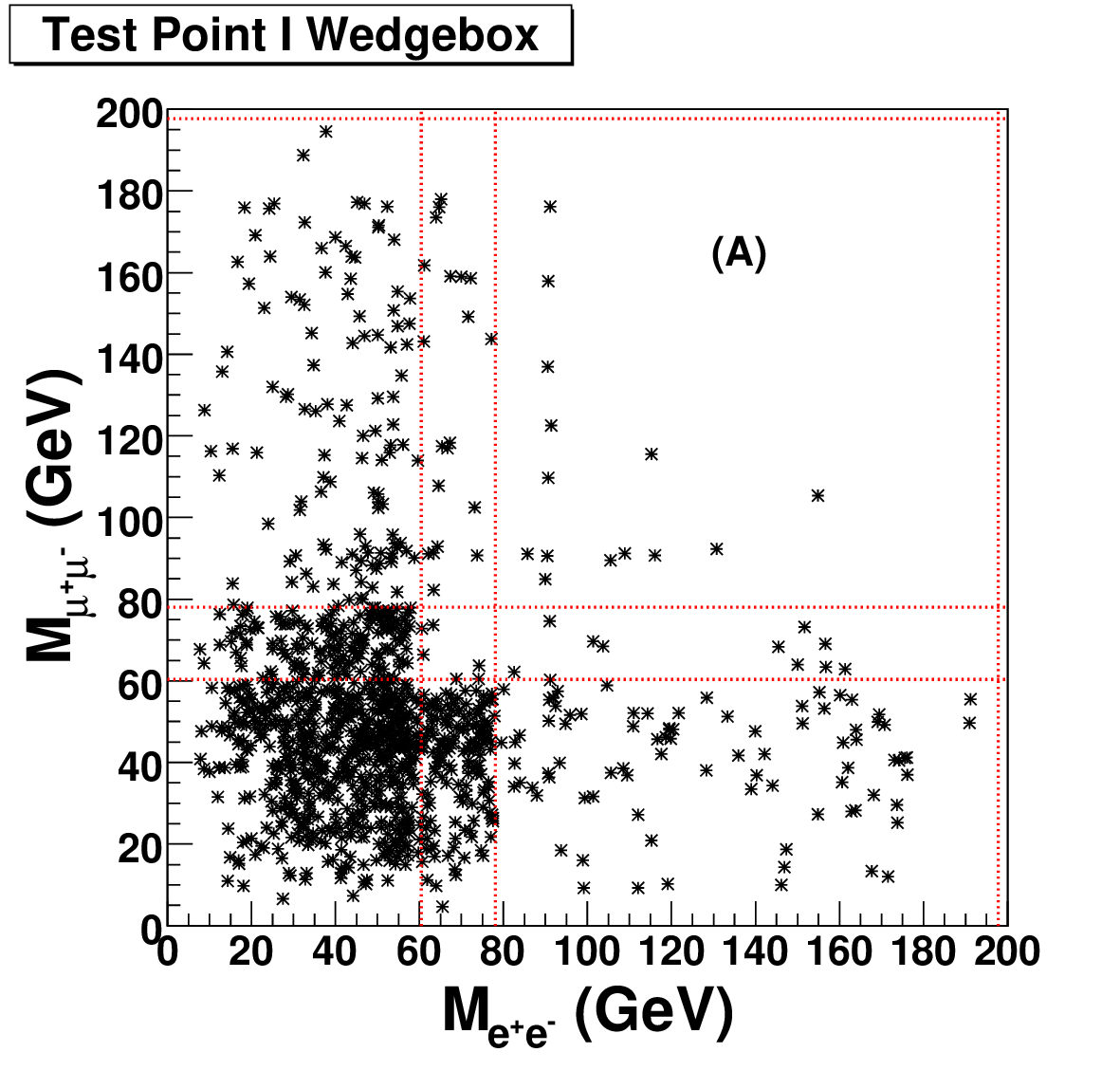}
        \includegraphics[width=1.8in]{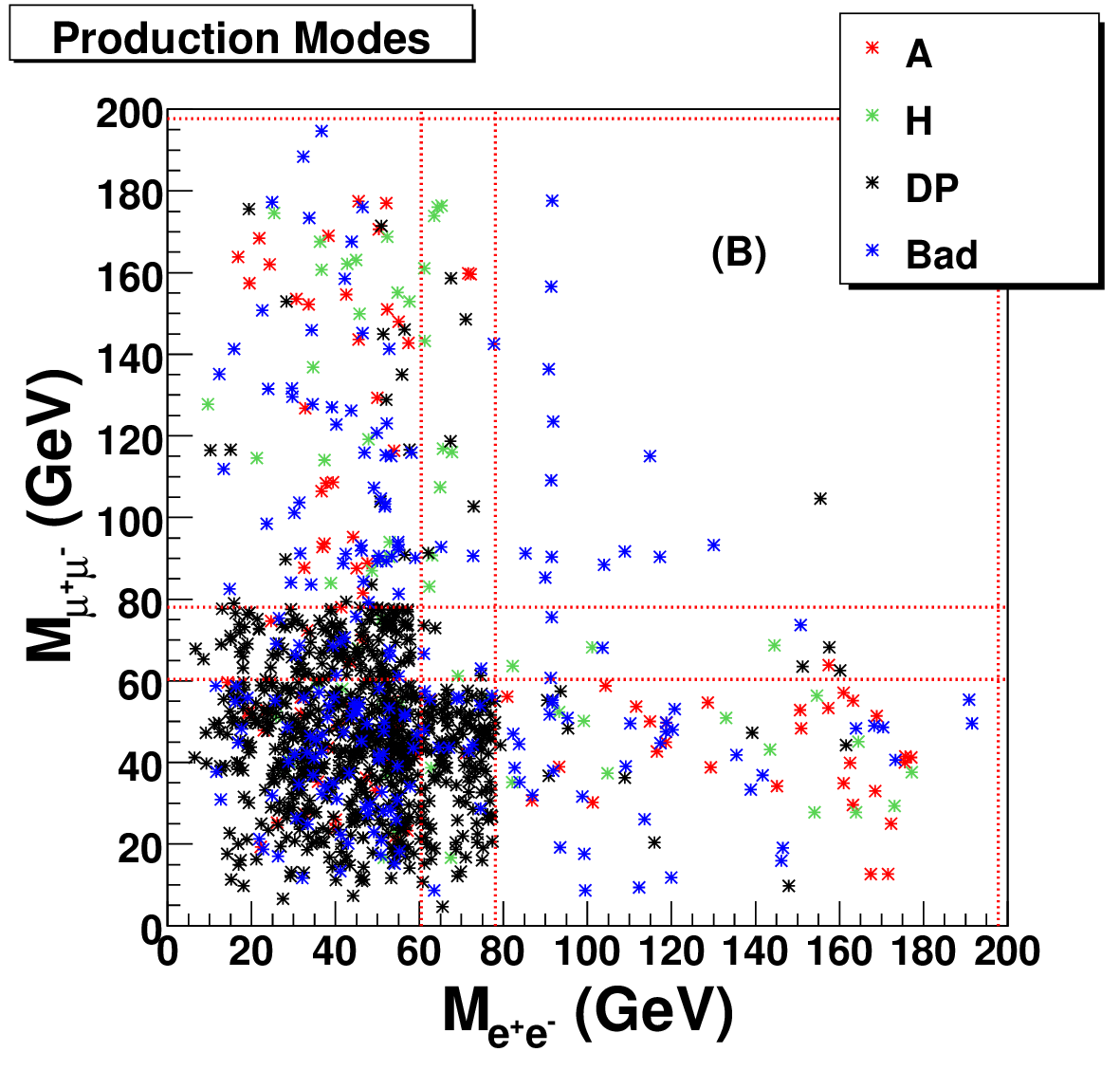}
        \includegraphics[width=1.8in]{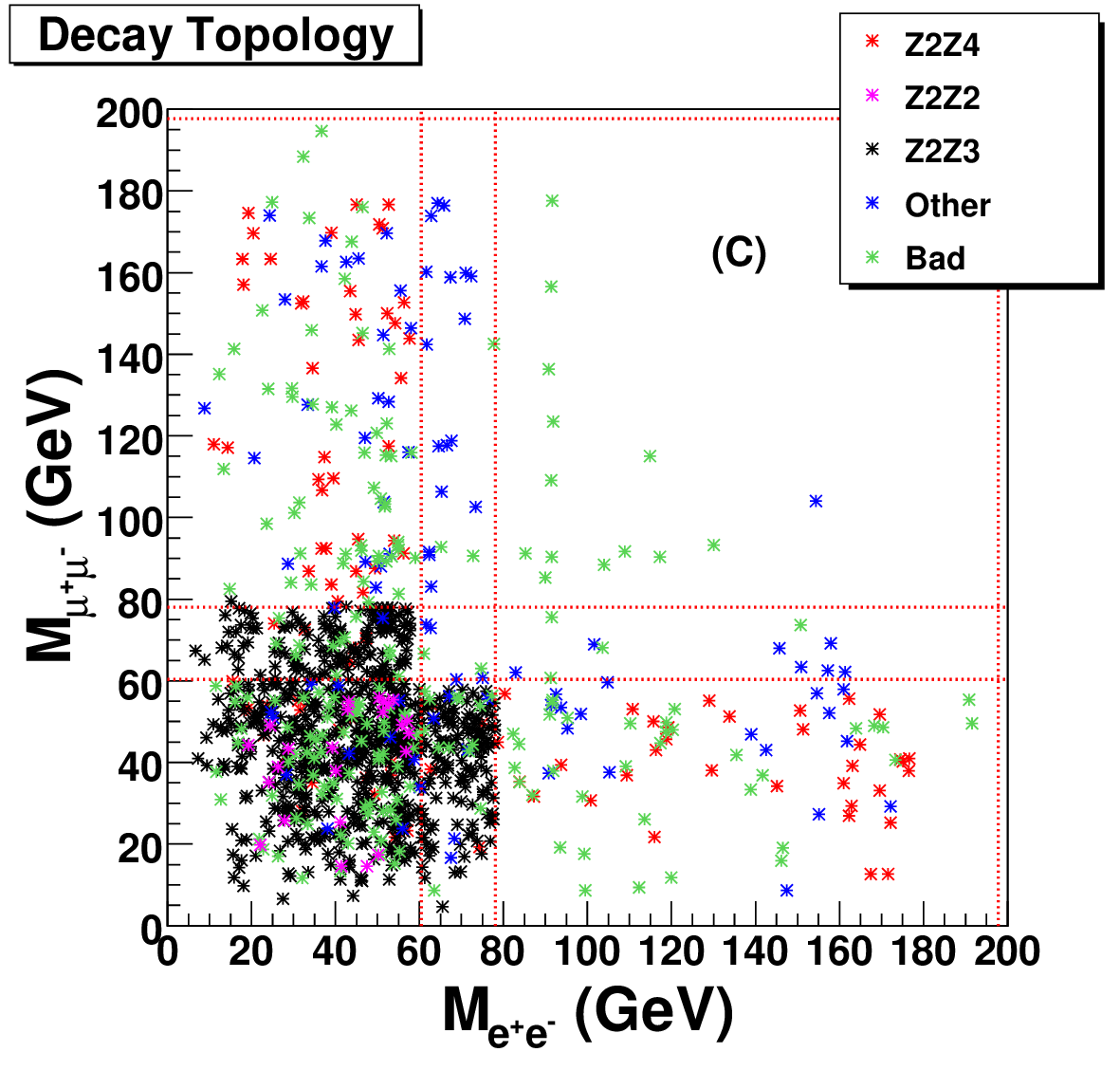}
    \end{center}
    \caption{\small \emph{ (A) Wedgebox plot for the
    {\bf MSSM Test Point I}
    defined in the text, for $300\, \hbox{fb}^{-1}$. (B) origin of each 
모모four-lepton event labeled as from `$A$', `$H$', `DP'(direct production), or
    otherwise `Bad'.
    (C) specific pair of neutralinos generated in each four-lepton
 event identified as  ${\widetilde\chi}^0_2 {\widetilde\chi}^0_4$ (`Z2Z4'),
 ${\widetilde\chi}^0_2 {\widetilde\chi}^0_3$ (`Z2Z3'),
 ${\widetilde\chi}^0_2 {\widetilde\chi}^0_2$  (`Z2Z2'),
   some other neutralino pair (`Other'),
   or other events `Bad'.}}
    \label{fig:wedgeplots}
\end{figure}

However, the shape of the wedgebox plot at this point suggests selecting events via their
decay topology:
Fig.~\ref{fig:wedgeplots}a shows a clear 'double wedge' topology implying that events due to
${\widetilde\chi}^0_2 {\widetilde\chi}^0_2$ decays are confined to the innermost box bounded by kinematic edges at $\sim 60\, \hbox{GeV}$, while those due to
${\widetilde\chi}^0_2 {\widetilde\chi}^0_3$ events are enclosed in the
legs of the short wedge terminating at $\sim 80\, \hbox{GeV}$.
Events from ${\widetilde\chi}^0_2 {\widetilde\chi}^0_4$ decays span both
of these regions and beyond up to the final kinematic edge at $\sim 200\, \hbox{GeV}$.

Strictly speaking, the structure of Fig.~\ref{fig:wedgeplots}a does not uniquely lead to the particle assignments noted in the previous paragraph.  In the MSSM, there are other processes that can generate edges in the wedgebox topology aside from those of the
form ${\widetilde\chi}^0_j \to l^+ l^- {\widetilde\chi}^0_1$, including
${\widetilde\chi}^\pm_2 \to l^+ l^- {\widetilde\chi}^\pm_1$ and
${\widetilde\chi}^0_j \to l^+ l^- {\widetilde\chi}^0_2$.  
The latter were given the appellation `stripes' in \cite{Bisset:2005rn} wherein they are discussed in some detail.
For instance, in principle, the shorter (longer) wedge could be due to 
${\widetilde\chi}^0_3 \to l^+ l^- {\widetilde\chi}^0_2$
(${\widetilde\chi}^0_3 \to l^+ l^- {\widetilde\chi}^0_1$) rather than due to 
${\widetilde\chi}^0_3 \to l^+ l^- {\widetilde\chi}^0_1$ 
(${\widetilde\chi}^0_4 \to l^+ l^- {\widetilde\chi}^0_1$) as described above.
Separate studies of decay kinematics can potentially exclude such alternative possibilities.  
For parameter set choices examined herein, these other feasible decay modes have a totally negligible effect.  More importantly, as will be discussed more in the next section, such ambiguities are largely irrelevant to the mass spectrum analyses described in this work. 

As in Fig.~\ref{fig:sps1a-full}(b,c), 
Fig.~\ref{fig:wedgeplots}(b,c) is color-coded\footnote{ Color-coding is of course an unfair advantage available in simulations but unavailable to experimentalists.
It is shown in the plots here to give the reader a clearer picture of what processes are actually occurring at significant rates.  Information from this color-coding
does not enter into any of the analyses results (mass values) found in this work.} 
to show the separate distributions of events from different production channels
(A, H, or 'direct' production $\equiv$ DP) and by their assorted neutralino-pair types; these distributions agree with remarks in the previous paragraph.
Note, however, the presence of `Bad' events which do not distribute
themselves nicely within the kinematic bounds and which are typically due
to chargino decays. Though nearly 10 percent of the total number of events
are `Bad몶 events, about half of these are excluded by rejecting events outside
the overall wedge structure, again nicely illustrating the strength of the
wedgebox technique\footnote{Recently, \cite{Bai:2010hd} has also discussed
ideas for excluding such bad events.}.
\begin{figure}[!thb]
\begin{center}
 \includegraphics[width=2.5in]{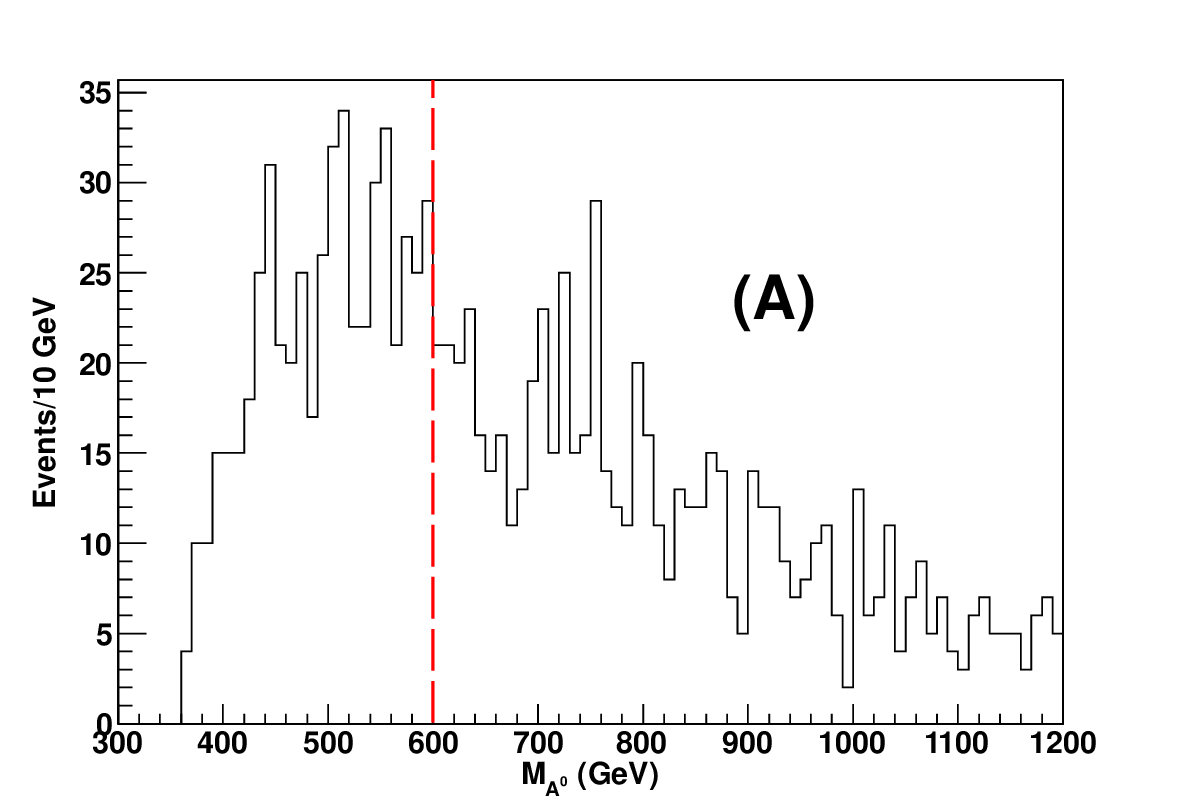}
 \includegraphics[width=2.5in]{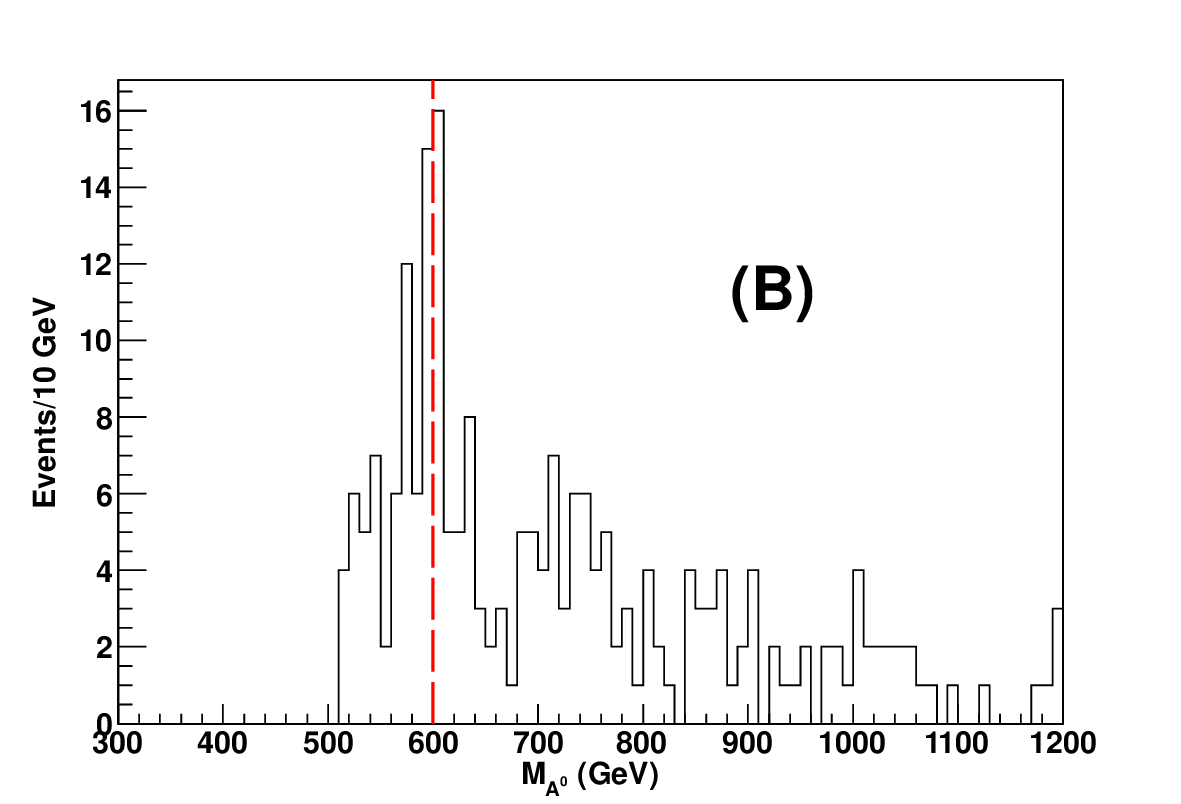}
\end{center}
\vskip -0.6cm
\caption{   \small \emph{(A) Distribution of solutions for events taken
from the inner box region of Fig. \ref{fig:wedgeplots}a; {\it i.e.}, within
the first set of dotted lines at  $ \sim 60\, \hbox{GeV}$. Here the correct
Higgs boson mass of $M_A \approx 600\, \hbox{GeV}$ does not appear at the
peak. (B) Distribution of solutions for events taken from the
outer legs of Fig. \ref{fig:wedgeplots}a; {\it i.e.}, within the regions
bounded by dotted lines at $\sim 80\, \hbox{GeV}$ and $\sim 200\, \hbox{GeV}$.
Here the correct Higgs boson mass of $M_A \approx 600\, \hbox{GeV}$
coincides with the peak.}
 \label{fig:histos} }
\end{figure}

Without the assistance of the wedgebox plot, one might be led to assume that
events with $M_{ee,\mu\mu} < 60\, \hbox{GeV}$ correspond mostly to the decay
topology of (\ref{Adecay}) with $i=j=2$.
This, however, leads to a Higgs boson mass distribution as shown in
Fig.~\ref{fig:histos}a.
There is neither a clear peak nor any kind of distinguishing feature near
the correct value of $M_A = 600\, \hbox{GeV}$.
Evidently, what might seem the natural choice of using events from the
densest region of the wedgebox plot is not optimal.

Instead, events should be selected from the most \emph{homogeneous} zone of
the wedgebox plot, which in this case consists of the outermost legs from
$80\, \hbox{GeV}$ to $200\, \hbox{GeV}$, corresponding mostly to
$A^0 \to {\widetilde\chi}^0_2 {\widetilde\chi}^0_4$ (even without looking
at Fig.~\ref{fig:histos}b, \cite{Bian:2006xx} predicts that events in
the outer wedge of a double-wedge plot come from Higgs boson decays).
As seen in Fig.~\ref{fig:histos}b, the N-MST method now works splendidly, yielding an easily-identified peak at the correct Higgs boson mass\footnote{Note from 
Fig.~\ref{fig:wedgeplots}b that the contributions from $A^0$ and from $H^0$ are
much more comparable than they were at SPS1a.  At {\bf MSSM Test Point I} the
mass difference between $H^0$ and $A^0$ is around $5\, \hbox{GeV}$ (as opposed
to under $3\, \hbox{GeV}$ at SPS1a), and their widths of $\sim \, 5\, \hbox{GeV}$ are about 5 times larger than at SPS1a (as determined by ISAJET and incorporated into this simulation).  These characteristics are reflected in the plot where the peak is spread over approximately two of the $5\, \hbox{GeV}$-wide bins.}, even though now an additional sparticle mass plays into the equations.  The goodness of fit is more surprising considering the number of 'bad' events distributed throughout this zone; inspection reveals that these latter, however, often fail to give solutions to equations (\ref{con1})-(\ref{con2}), so they do not heavily interfere with the shape of the distribution.

Improvement of the N-MST method via a wedgebox plot, at least for the
Higgs boson decay topology considered here, is therefore quite straightforward.
However, these improvements may not always be able to counter some of the
short-comings of the N-MST model, namely, the assumptions of the known sparticle masses and zero net transverse momentum --- Eqns. (\ref{con2}).  The sparticle masses are assumed to be determined via some unspecified preceding analysis, and it would be more correct to attach uncertainties to these values rather than just input the exact values given by the simulator at this point in the parameter space. It is also quite possible that the mass of the heavier neutralino $\widetilde{\chi}^0_{4}$ required for the analysis at {\bf MSSM Test Point I} may be far less accurately determined (or left unknown) by said nameless preceding analyses than are the masses of the lighter sparticles ($\widetilde{\chi}^0_{1}$, $\widetilde{\chi}^0_{2}$, and $\tilde{\ell}^{\pm}$) which suffice at the SPS1a point.  Thus a study in which both the Higgs boson mass and the mass of the heavier neutralino are simultaneously determined would certainly have merit.  Better still would be to jettison reliance on such un-named previous studies and to determine all the to-date unknown beyond-the-standard-model particle masses in a self-contained analysis.   This is the aim of the C-MST method to which this paper now turns (as opposed to piling more details into a fundamentally-weaker N-MST analysis).  The slightly more pedagogical goal of this N-MST section has been to succinctly demonstrate the worth of a combined MST \& wedgebox analysis without immediately introducing numerous subtle (and potentially distracting) issues inherent in an even more realistic and self-contained study.  

\begin{figure}[htpb]
    \begin{center}
        \includegraphics{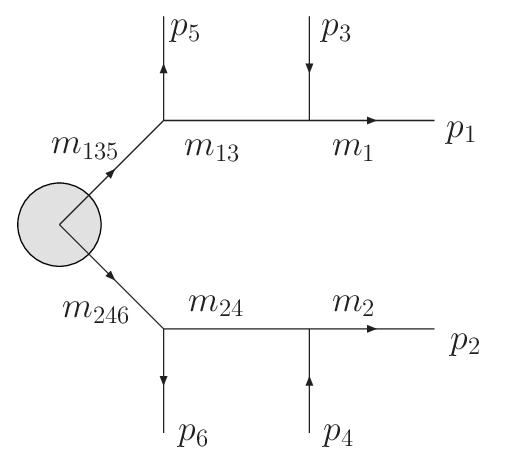}
    \end{center}
    \caption{Event topology, taken from \cite{Cheng:2007xv,Cheng:2008mg,Cheng:2009fw}.}
    \label{fig:event}
\end{figure}

\section{The C-MST Method of Cheng {\it et al.}}
\label{sec:cheng}
Next consider the C-MST method.
The process treated by Cheng {\it et al.} \cite{Cheng:2007xv,Cheng:2008mg,Cheng:2009fw}
is illustrated in Fig.~\ref{fig:event}
in which the  masses $m_1$,  $m_2$, $m_{13}$,  $m_{24}$, $m_{135}$ and
$m_{246}$ must satisfy a set of equations precisely analogous to
(\ref{con1})-(\ref{con2}):
\begin{eqnarray}
\label{guneq1}
   m_1^2  &=& p_1^2 \\
   m_2^2  &=&  p_2^2 \\ \label{guneq3}
   m_{13}^2 &=& \left( p_1 + p_3 \right)^2  \\ \label{guneq4}
   m_{24}^2 &=& \left( p_2 + p_4 \right)^2 \\
   m_{135}^2 &=&\left( p_1 + p_3 + p_5 \right)^2 \\
   m_{246}^2 &=& \left( p_2 + p_4 + p_6
    \right)^2 \\
  p_\text{sum}^x &=& \sum_i p^x_i \\
  p^y_\text{sum} &=&  \sum_i p^y_i
  \label{guneq8}
\end{eqnarray}
where the transverse momentum sums $p_\text{sum}^{x,y}$ are assumed to be
calculable from measurements of associated jet momenta (produced, though
not shown, in the gray bubble in Fig.~\ref{fig:event}) and missing momentum
(from the LSPs) necessary to balance the whole:
\begin{eqnarray}
  p^x_\text{sum} = \sum_l p^x_l  + p^x_\text{miss}
  \qquad p^y_\text{sum} = \sum_l p^y_l + p^y_\text{miss}
    \label{eqn:gunpt}
\end{eqnarray}

The specific example considered in \cite{Cheng:2007xv,Cheng:2008mg,Cheng:2009fw} was production of two
neutralinos via squarks in the MSSM, followed by decay
via on-shell smuons to muons and two LSPs:
\begin{equation}\label{neutdecay}
  \tilde{q} \tilde{q} \to q ~q ~\widetilde{\chi}_{2}^0~
   \widetilde{\chi}_{2}^0 (\to \tilde{\mu}^\pm {\mu}^\mp
   \to  \mu^+ \mu^-  \widetilde{\chi}_{1}^0)
\end{equation}
giving
\begin{align}
    m_1 &= m_2 = m_{\tilde{\chi}^0_1} \nonumber \\
    m_{13} &= m_{24} = m_{\tilde{\mu}} \nonumber \\
    m_{135} &= m_{246} = m_{\tilde{\chi}^0_2}
    \label{eqn:mass_definition}
\end{align}
In the three-dimensional space of masses
$(m_{\widetilde{\chi}^0_1}, m_{\widetilde{\chi}^0_2}, m_{\widetilde{\mu}})$,
each event gives eight equations (\ref{guneq1})-(\ref{guneq8}) for the eight
unknown LSP momenta $p_1^\mu$ and $p_2^\mu$, assuming the outgoing muon
momenta $p_3, p_4, p_5, p_6$ can be measured.
The solution to this set of equations is again, as for the system of
(\ref{con1})-(\ref{con2}), a quartic equation with 0, 2, or 4 real roots.
In contrast to the discussion of the last section, however, rather than trying
to find the point in
$(m_{\widetilde{\chi}^0_1}, m_{\widetilde{\chi}^0_2},
m_{\widetilde{\mu}})$-space
where the density of solutions ($\equiv \mathbb{N}
(m_{\widetilde{\chi}^0_1}, m_{\widetilde{\chi}^0_2}, m_{\widetilde{\mu}})$)
is maximized,
instead the point where the \emph{gradient} of $\mathbb{N}$ is maximized
(a heuristic argument for this is given in \cite{Cheng:2007xv,Cheng:2008mg,Cheng:2009fw}) is sought.

In \cite{Cheng:2007xv,Cheng:2008mg,Cheng:2009fw} this method apparently works quite well at the MSSM parameter points studied, giving all relevant sparticle masses to a few percent after collection of data samples corresponding to 
$300\, \hbox{fb}^{-1}$ of integrated LHC luminosity.
However, to test the robustness of this technique and the extent to which
augmentation with the wedgebox technique can improve results, consider a new 
MSSM parameter point (see Tab.~\ref{tab:masses} again for masses):
\begin{description}
\item[MSSM Test Point II]
\begin{eqnarray*}
    \mu = -150\, \hbox{GeV}  ~~ ~~~~ M_2 = 250\, \hbox{GeV} ~~~~ ~~
    M_1 = 90\, \hbox{GeV}\\
    \tan \beta = 5 ~ ~~~~
    M_{\widetilde{e, \mu}_L, \widetilde{\tau}} = 250\, \hbox{GeV} ~ ~~~~
    M_{\widetilde{e, \mu}_{R}} = 120\, \hbox{GeV}
      \\
    M_A = 700\, \hbox{GeV} ~~~~~ M_{\tilde{q}} = 400\, \hbox{GeV} ~~~~
    M_{\tilde{g}} = 500\, \hbox{GeV}
\end{eqnarray*}
\end{description}

\begin{figure}[!thb]
\begin{center}
\dofig{4.15in}{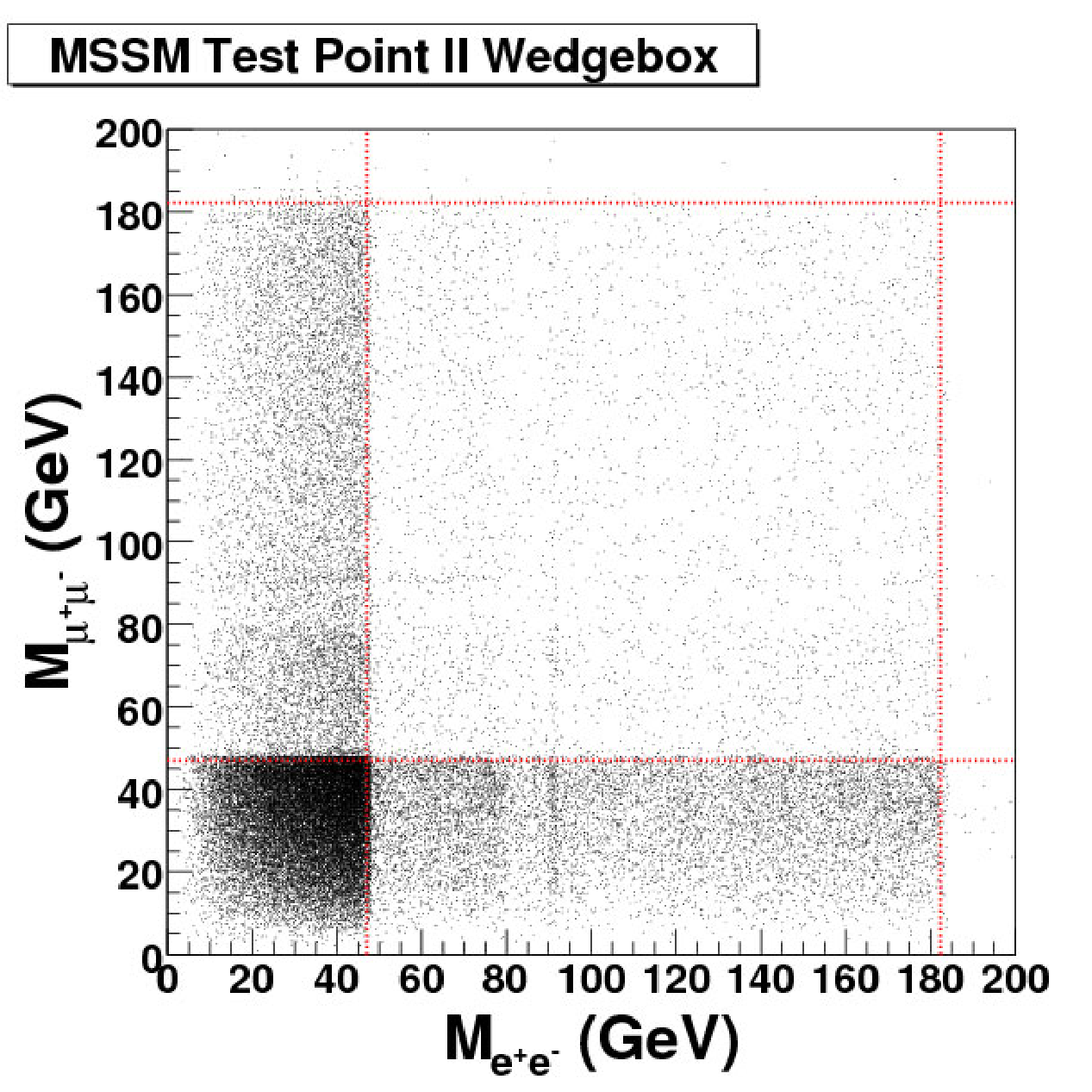}
\end{center}
\vskip -0.6cm
\caption{ \small \emph{Wedgebox at {\bf MSSM Test Point II} for a luminosity
of $90\, \hbox{fb}^{-1}$. }
 \label{t2-wedge} }
\end{figure}

\subsection{Wedgebox selection}

The wedgebox structure\footnote{ {\bf MSSM Test Point II} is clearly
representative of the general case in the MSSM, where $pp$ collisions
yield a `mixed bag' of concomitant neutralino decays.
Events on this plot pass the same cuts as in Section \ref{sec:add},
minus the jet cut.} of Fig.~\ref{t2-wedge}
is again generated using ISAJET
\cite{Baer:1999sp,Paige:2003mg} and the event selection criteria mentioned earlier
(save no jet cut); the plot compartmentalized into four substantially 
event-populated regions by the shown red-dashed lines to which will 
be applied the following nomenclature:
\newline
The Zone 1 box, with
$[M_{ee}\, \hbox{and} \, M_{\mu\mu}] < 47\, \hbox{GeV}$,
is the most densely-populated region of the wedgebox plot
and should include all
${\widetilde{\chi}^0_2} {\widetilde{\chi}^0_2}$ events.
\newline
Zone 2 is composed of two rectangles (the legs of wedges) running outwards
along both axes from the Zone 1 box --- satisfying the condition that
$[M_{ee}\, \hbox{and} \, M_{\mu\mu}] < 182\, \hbox{GeV}$ \emph{and}
$[M_{ee}\, \hbox{or}\, M_{\mu\mu} \, \hbox{but not both}] < 47\, \hbox{GeV}$.
Events due to ${\widetilde{\chi}^0_2} {\widetilde{\chi}^0_3}$ and
${\widetilde{\chi}^0_2} {\widetilde{\chi}^0_4}$ not residing in
Zone 1 will fall\footnote{With the
${\widetilde{\chi}^0_2} {\widetilde{\chi}^0_3}$
(${\widetilde{\chi}^0_2} {\widetilde{\chi}^0_4}$)
events terminating at
$M_{ee}~\hbox{or}~M_{\mu\mu} \simeq 80\, \hbox{GeV}$
($182\, \hbox{GeV}$).  The endlines at $\simeq 80\, \hbox{GeV}$
are faintly discernible in Fig.~\ref{t2-wedge}; however, the forthcoming
analysis does not rely upon this.} in Zone 2.
\newline
Zone 3, with
$47\, \hbox{GeV}< [M_{ee}~and~M_{\mu\mu}] < 182\, \hbox{GeV}$, lies outside
of Zones 1 \& 2 and should only be populated by
${\widetilde{\chi}^0_3} {\widetilde{\chi}^0_3}$,
${\widetilde{\chi}^0_3} {\widetilde{\chi}^0_4}$
and
${\widetilde{\chi}^0_4} {\widetilde{\chi}^0_4}$
events.

\begin{figure}[!thb]
\begin{center}
\dofig{4.15in}{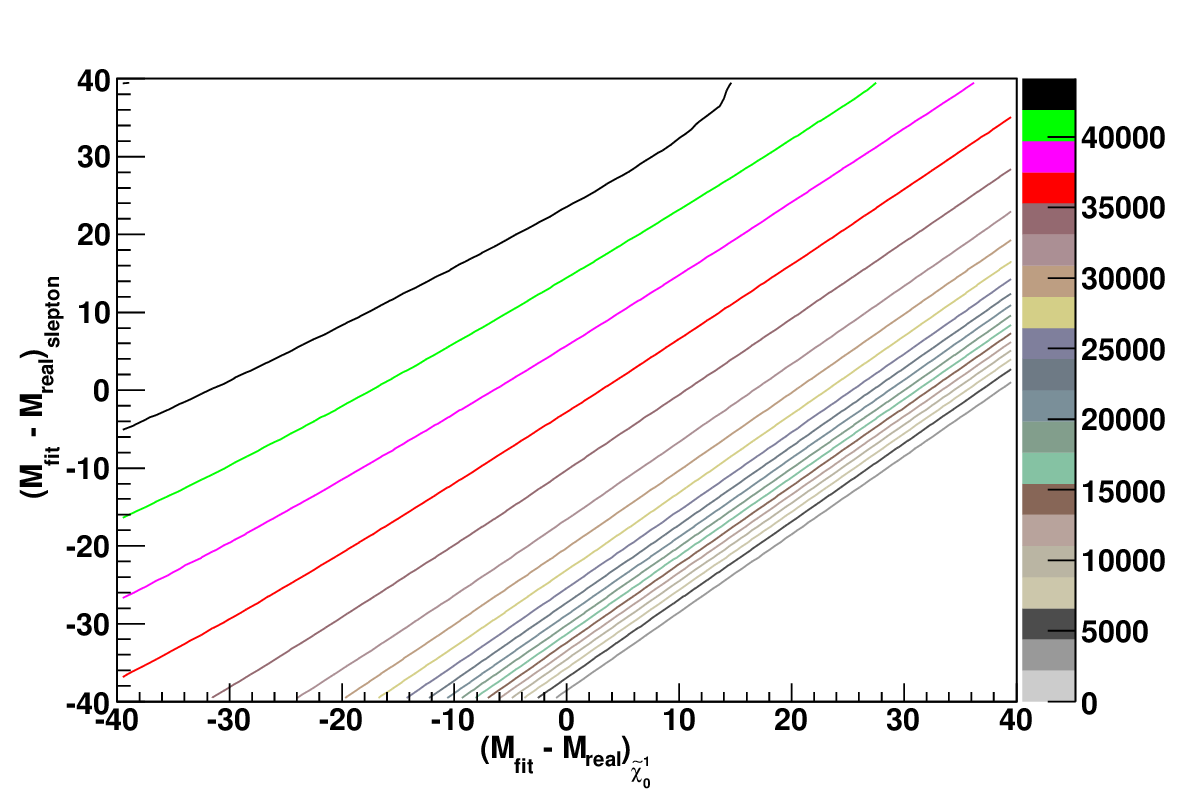}
\end{center}
\vskip -0.6cm
\caption{ \small \emph{Two-dimensional mass space scan for MSSM Test Point II, 
seeking the point of steepest descent via the C-MST method, using the whole
sample from the wedgebox plot. The axes are the difference between the input
mass and the real mass of LSP and slepton.}
 \label{fig:badfit} }
 \end{figure}

Consider first using four-lepton events across the whole wedgebox plot 
--- this would include several different kinds of neutralino pair events. 
A scan was performed over the $(m_{\widetilde{\chi}^0_1},
m_{\widetilde{\chi}^0_2},
m_{\widetilde{\mu}})$ mass space, and the resulting values projected onto the
$(m_{\widetilde{\chi}^0_1}, m_{\widetilde{\mu}})$-plane. As Table 1 shows
$m_{\widetilde{\mu}} - m_{\widetilde{\chi}^0_1} \sim 41 \hbox{GeV}$,
Fig.~\ref{fig:badfit} indicates that the gradient of $\mathbb{N}$ maximizes along the line
$m_{\widetilde{\chi}^0_1}= m_{\widetilde{\mu}}$ rather than at the desired
true value: the C-MST method fails \emph{badly} in this attempt to find the
sparticle masses.  Since the wedgebox plot clearly indicates that
${\widetilde{\chi}^0_2} {\widetilde{\chi}^0_3}$,
${\widetilde{\chi}^0_2} {\widetilde{\chi}^0_4}$
and
${\widetilde{\chi}^0_4} {\widetilde{\chi}^0_4}$ production is
substantial, the failure of the C-MST is not surprising given the
inhomogeneity of the data set.

Consider instead performing an analysis limited to events in Zone 3 --- 
which should be more homogeneous.  Before proceeding though one more feature
of the wedgebox plot should be taken into account:  there is a
clear $Z$-line at
$[M_{ee}\, \hbox{or} \, M_{\mu\mu}] \simeq 91\, \hbox{GeV}$.
This is due to neutralinos decaying to an LSP and a pair of leptons via
an on-shell $Z^0$ rather than via a slepton\footnote{Although the
missing energy cut should eliminate most SM $Z^0Z^0$, $Z^0Z^{0*}$ events,
any such remnant background surviving would also populate this line.}.
The situation is greatly simplified if the events in Zone 3 are further
curtailed to encompass only those for which
$100\, \hbox{GeV}< [M_{ee}\, \hbox{and} \, M_{\mu\mu}] < 182\, \hbox{GeV}$.
This truncated Zone 3 will be called Zone 3$^{\prime}$.
This will exclude the on-shell $Z^0$ events along with those due to
${\widetilde{\chi}^0_3} {\widetilde{\chi}^0_3}$
or
${\widetilde{\chi}^0_3} {\widetilde{\chi}^0_4}$.
The remaining fairly homogeneous subset of events still yields 1000+ signal
events (corresponding to $90\, \hbox{fb}^{-1}$ of LHC integrated luminosity),
80\% of which are in fact due to
${\tilde{\chi}^0_4} {\tilde{\chi}^0_4}$ pair production
(the remainder mostly involve colored sparticle decays into the heavier
chargino, $\widetilde{\chi}_2^{\pm}$).

At this juncture it is appropriate to revisit a statement made in the previous section --- that ambiguities in identifying the particle identities from the wedgebox plot
topology are largely irrelevant to the analysis to be performed.  As noted in the case of {\bf MSSM Test Point I}, the occurrence of `stripes몶 could 
mean that zones of the wedgebox plot should be re-assigned.  For instance, 
Zone 3 could be due to 
${\widetilde\chi}^0_{3} \to {\widetilde\chi}^0_1$ decays
(and thus more correctly referred to as a 
${\widetilde{\chi}^0_3} {\widetilde{\chi}^0_3}$ region), 
while Zone 2 has many `stripe몶 events 
--- ${\widetilde\chi}^0_{3} \to {\widetilde\chi}^0_2$ decays.
{\em This is {\bf not} true at this point in the parameter space.} 
Since the goal is to select the region of the wedgebox plot with the purest 
event set, the outer-most (up and to the right) clearly delineated region
is chosen to perform the analysis. Said region will never be due to 몵stripe몶
events --- the isolated process will be of the form  
${\widetilde\chi}^0_j \to l^+ l^- {\widetilde\chi}^0_1$, where $j = 3$ or $4$.  
The mass value obtained from the analysis will be correct, the only question 
will be whether it is $m_{{\widetilde\chi}^0_{3}}$ or $m_{{\widetilde\chi}^0_{4}}$.  Further input may be required to resolve this uncertainty.  
Likewise, if the wedgebox plot topology was a single wedge\footnote{If the wedgebox plot is a single box, then this is virtually certainly ${\widetilde{\chi}^0_2} {\widetilde{\chi}^0_2}$, even if in principle one could attribute it to 
${\widetilde{\chi}^0_3} {\widetilde{\chi}^0_3}$ or 
${\widetilde{\chi}^0_4} {\widetilde{\chi}^0_4}$ 
with all other rates somehow suppressed.}, then the legs of the wedge 
could be due to ${\widetilde{\chi}^0_2} {\widetilde{\chi}^0_3}$ or 
${\widetilde{\chi}^0_2} {\widetilde{\chi}^0_4}$
(in this latter case presumably something in the coupling heavily suppresses the rate for
${\widetilde\chi}^0_3 \to l^+ l^- {\widetilde\chi}^0_1$, and so such events make a negligible impact on the wedgebox plot).  Again, an analysis of the type presented herein would yield a
correct mass determination, only with some ambiguity as to the naming of the particle involved.   

Additional improvements in the numerical results are possible if another piece of 
information is utilized:
the actual value\footnote{It is acknowledged that this runs counter to 
the statement in the introductory remarks that the edge locations are used 
purely to delineate the data set and not to provide additional numerical 
information. However, it would be foolish to completely ignore this extra 
information that is readily accessible.  The aim herein is to emphasize how 
the wedgebox can be used to purify a data set, not to exclude the use of 
additional useful information.}  for the edge of the 
$\widetilde{\chi}^0_4\widetilde{\chi}^0_4$ box, $\Delta$.
It is clear that this edge can be measured quite precisely
via the wedgebox or the traditional one-dimensional triangular mass distribution, yielding a relation between the three masses in the equation:
\begin{equation}
    \Delta =
    m_{\widetilde{\chi}^0_4}\sqrt{1-\frac{m_{\widetilde{\mu}}^2}
    {m_{\widetilde{\chi}^0_4}}^2}\sqrt{1-\frac{m_{\widetilde{\chi}^0_1}^2}
    {m_{\widetilde{\mu}}^2}} \;\;\; .
    \label{eqn:mass}
\end{equation}
This additional information enables one to scan for only
$m_{\widetilde{\chi}^0_1}$ and $m_{\widetilde{\mu}}$, with the
$\widetilde{\chi}^0_4$ mass calculated by solving the equation above.
\par

\begin{figure}[!thb]
\begin{center}
\dofig{4.15in}{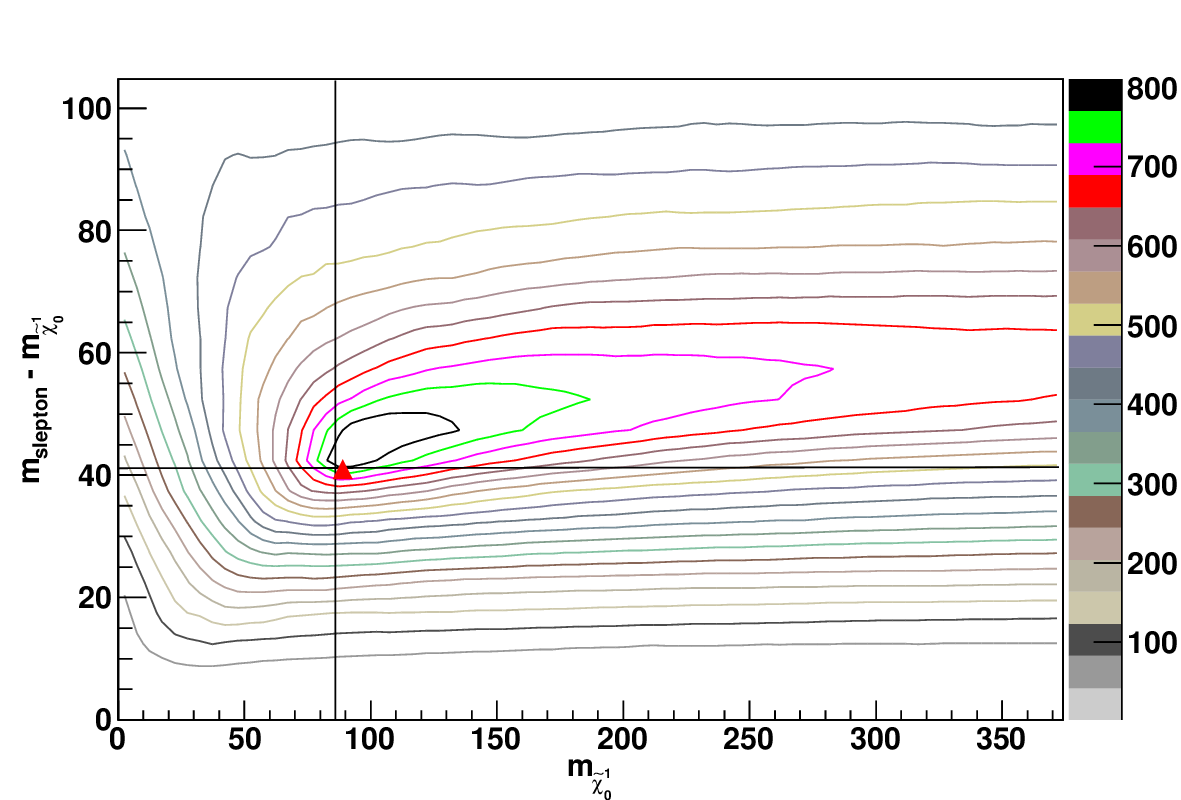}
\end{center}
\vskip -0.6cm
\caption{ \small \emph{Two-dimensional scan result 
seeking the point of steepest descent via the C-MST method,
using events from Zone 3$^{\prime}$ without detector effects and with correct lepton 
placements.
The horizontal axis is the $\widetilde{\chi}^0_1$ mass, and the vertical axis
is the mass difference of slepton and $\widetilde{\chi}^0_1$} The horizontal
and vertical lines show the correct values for $m_{\widetilde{\mu}} -
m_{\widetilde{\chi}^0_1}$ and $m_{\widetilde{\chi}^0_1}$, respectively.
The triangle gives the optimal location obtained from the C-MST steepest
descent analysis.
 \label{fig:theoretical} }
\end{figure}

Proceeding somewhat gradually, first consider applying an analysis 
incorporating $\Delta$ to the neutralino events from Zone 3$^{\prime}$, but 
with no detector effects and with the simulation co-opted to only include
correct lepton placements in Fig.~\ref{fig:event}. The result, shown in 
Fig.~\ref{fig:theoretical}, indicates that the desired physical mass is 
indeed given quite accurately by the point of steepest descent for 
$\mathbb{N}$.

\begin{figure}[!thb]
\begin{center}
\dofig{4.15in}{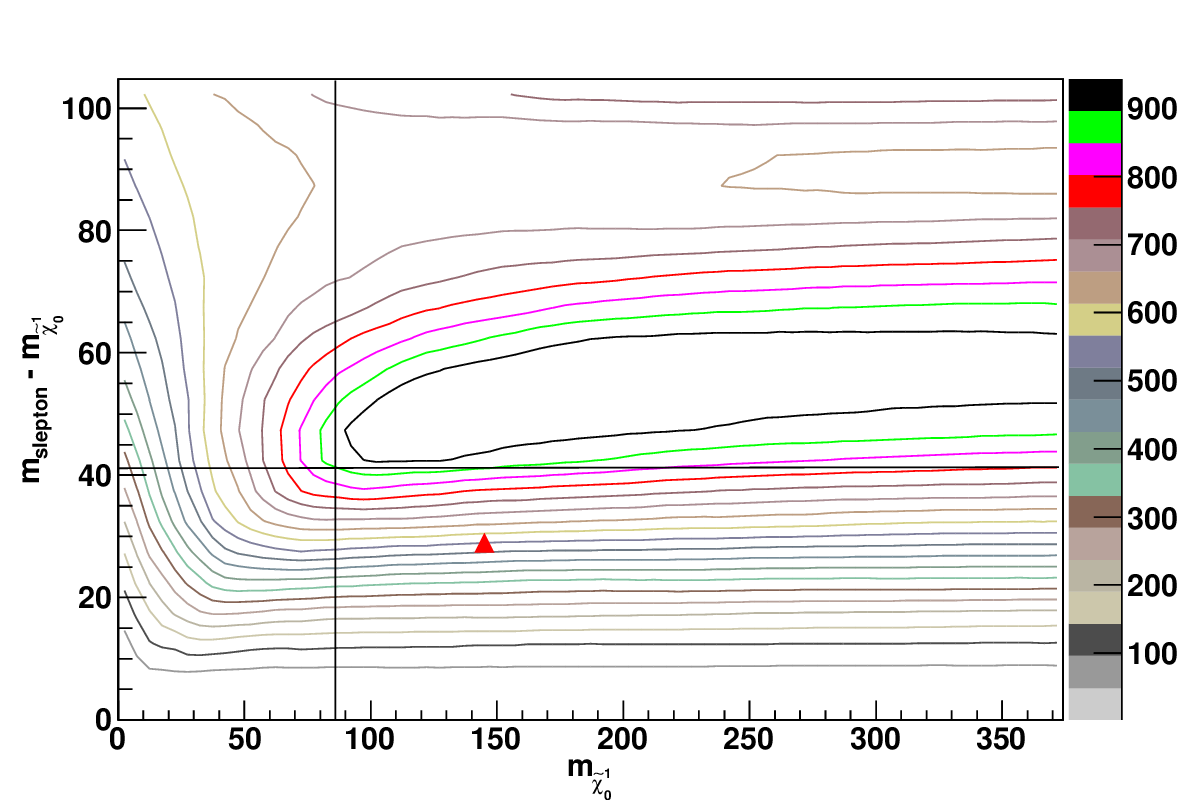}
\end{center}
\vskip -0.6cm
\caption{ \small \emph{Two-dimensional scan result 
seeking the point of steepest descent via the C-MST method,
using 1080 events from Zone 3$^{\prime}$ and also incorporating detector effects 
and excluding simulation information not available to the experimentalists concerning correct lepton placement.
The triangle gives the optimal location obtained from the C-MST steepest
descent analysis.
}
 \label{fig:goodfit} }
\end{figure}

Continuing on to the more realistic Zone 3$^{\prime}$ analysis including
detector smearing and possible lepton placement errors leads to the result
shown in Fig.~\ref{fig:goodfit}.  Clearly results are worse than in 
Fig.~\ref{fig:theoretical}: the point of the steepest descent of
$\mathbb{N}$ is still in the neighborhood of the physical mass
(there is in fact a local maximum quite close to the correct value, but it is not the global maximum indicated by the triangle), but it is obviously not a sharp maximum in the solutions space.   

\begin{figure}[!thb]
\begin{center}
\dofig{4.15in}{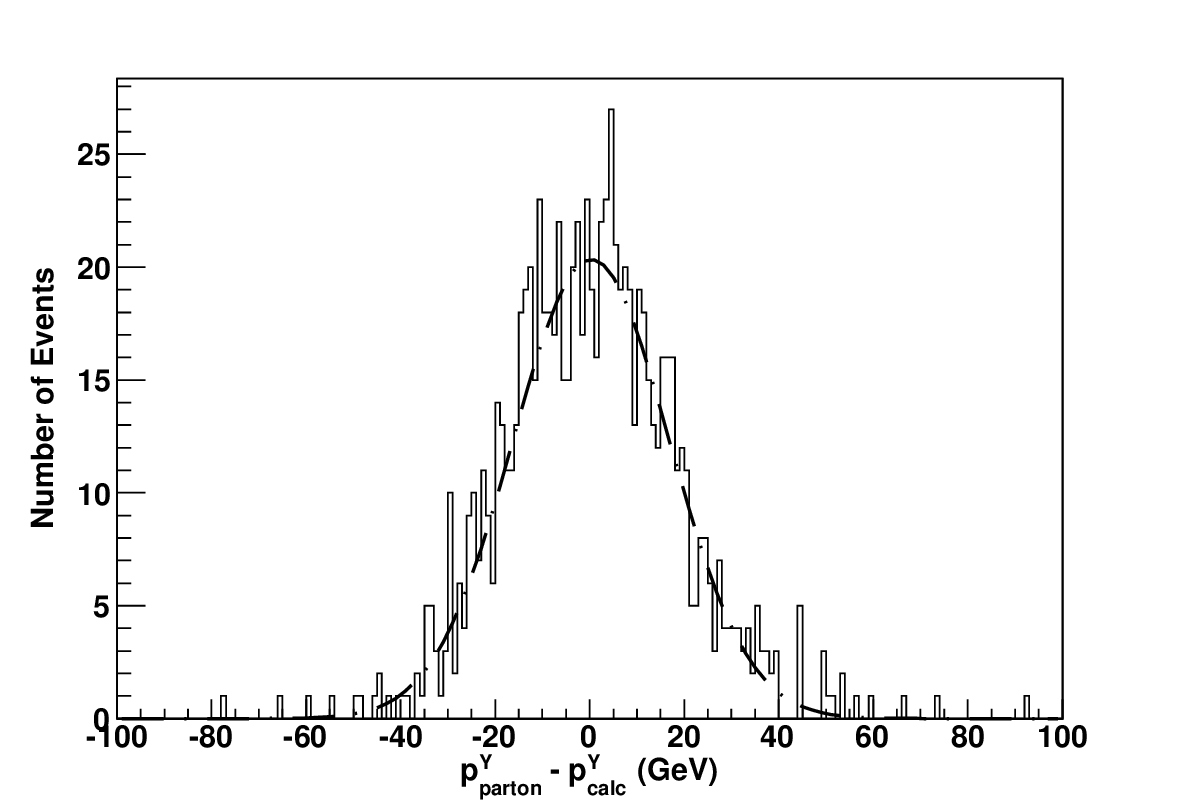}
\end{center}
\vskip -0.6cm
\caption{ \small \emph{$p_T$ distribution for one of the transverse
directions. Here $p_{parton}$ designated the net momentum of the two 
neutralinos (prior to their decay) while $p_{calc}$ is the sum of the 
four leptons' momenta plus the calculated missing momentum.  Thus
$p_{calc}$ includes detector smearing effects. 
Superimposed as a dashed curve on the simulation data is a Gaussian fit
whose standard deviation is $16.5\, \hbox{GeV}$ (and with a mean value of
$0.3\, \hbox{GeV}$)}
 \label{pt} }
 \end{figure}

\begin{figure}[htpb]
   \begin{center}
       \includegraphics[width=3.5in]{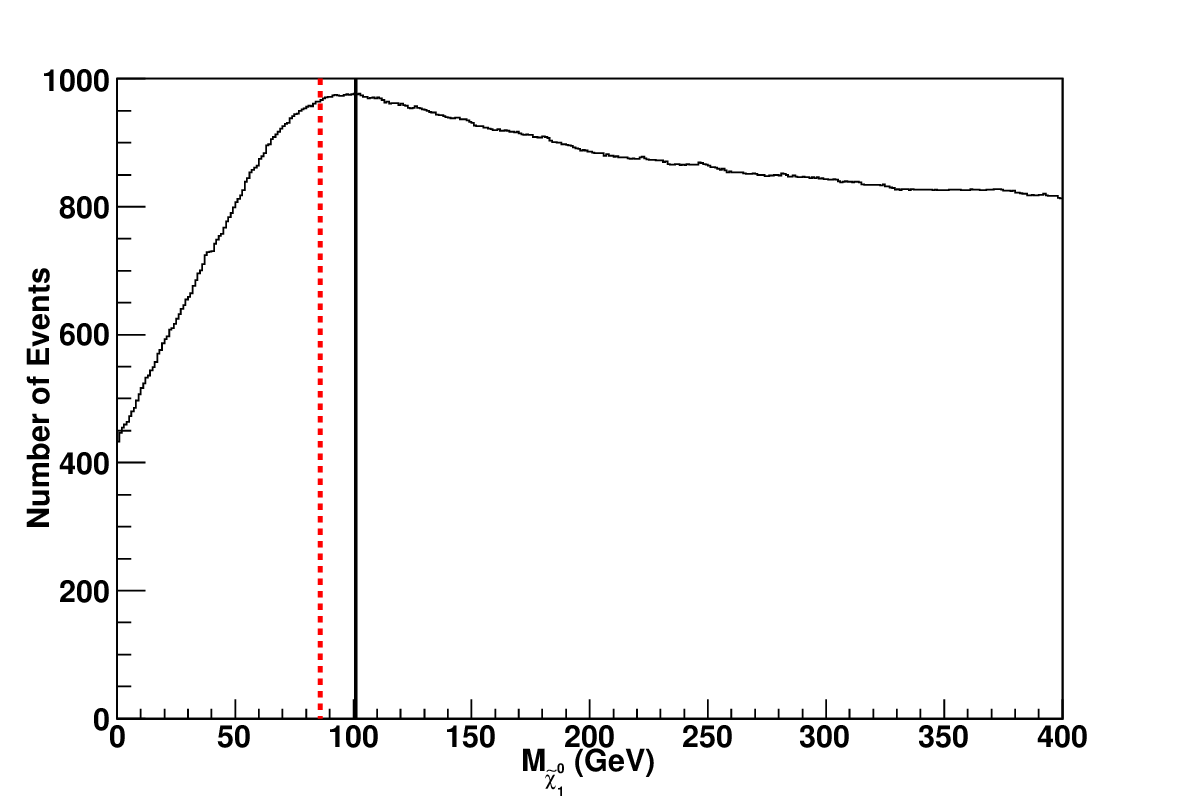}
   \end{center}
   \caption{\small \emph{1-D projection of Fig.~\ref{fig:goodfit}, assuming
      the correct value of $m_{\widetilde{\mu}}-  m_{\widetilde{\chi}^0_1}$,
      and scanning over $-16\, \hbox{GeV} < p_T < 16\, \hbox{GeV}$.
      The maximum of the curve,
      $m_{\widetilde{\chi}^0_1} \, \simeq \, 100 \, \hbox{GeV}$,
      roughly approximates the actual LSP mass ($86\, \hbox{GeV}$), given by
      the red dashed line. }}
    \label{fig:finalfit}
\end{figure}

One reason for persisting inaccuracy is difficulty in applying the
momentum conservation relation (\ref{eqn:gunpt}) --- primarily due to 
detector smearing effects.  With a simulation, one is able to compare
the ('parton level') net transverse momentum of the two neutralinos 
produced by the generator (prior to their decay), labeled as $p_{parton}$,
to the sum of the momenta of the four leptons in the signal plus the 
missing momentum, designated as $p_{calc}$.  
If $p_{parton}$ and $p_{calc}$ actually match the quantities on the 
left- and right-hand sides of Eqns.\ (\ref{eqn:gunpt}), respectively, 
then they should be equal.  However, as illustrated by Fig.~\ref{pt}, 
this is not the case.  The difference arises from the detector 
smearing\footnote{Particle resolutions are as given in a previous footnote.}
of the lepton momenta (in addition the smearing of the momenta of the 
other observed particles alters the value calculated for the missing 
momentum\footnote{Another possible source of error is if the missing
momentum is not solely due to the two LSP's produced in the neutralino
decays.  There could also be SM neutrinos produced in some events.  With 
cascade production, the initial gluinos or squarks will lead to quark jets
with significant $p_T$ in addition to the desired neutralino pair.  Decays
of heavy-flavored quarks within these quark jets may well yield such 
neutrinos, especially if heavy-flavored sparticle production or decays
of gluinos into heavy-flavored quarks is enhanced, as is expected in some 
scenarios \cite{Kadala:2008uy}. How such neutrinos might affect a CMST-style
analysis is currently under investigation \cite{WIP1}.}).
The range of the imbalance between the experimentally-measurable value
of $p_{calc}$ and the desired value of $p_{parton}$ is 
$-30\, \hbox{GeV} ~\gsim~ p_T ~\gsim ~30\, \hbox{GeV}$. 
This is formidable in light of the fact
that a small inaccuracy in $p_T$ (say $ 2\, \hbox{GeV}$) can change
the number of solutions at a given mass point by 10\%.
If smearing effects could be eliminated, so that $p_{calc}$ would 
essentially be equal to $p_{parton}$, fits to masses would be quite good 
(easily within 1\%).  Unfortunately, actual experiments cannot know 
how large the $p_T$ imbalance between $p_{parton}$ and $p_{calc}$ is in any
given event:  the best that can be done is to perform
a scan over this $p_T$ uncertainty for each event, taking for instance
$-16\, \hbox{GeV} < p_T < 16\, \hbox{GeV}$. This is done in making
Fig.~\ref{fig:finalfit}, whose maximum does lie somewhat closer
to the actual LSP mass; however, since the correct value of
$m_{\widetilde{\mu}} - m_{\widetilde{\chi}^0_1}$ was assumed in this
1-D projection, the result is somewhat better than what would be obtained 
in practice.  The fitted LSP mass in Fig.~\ref{fig:finalfit} has a small 
error but with a slightly up-biased central value (primarily due to the fact 
that the neutralino pairs do not have a fixed CM energy).
 
\section{Discussion \& Conclusions}
\label{sec:conc}

Use of a wedgebox plot to select the most homogeneous sample of events
has been shown to increase the accuracy and efficacy of
the N-MST and C-MST mass reconstruction methods.
If the events analyzed do not mostly share the same decay topology, 
both methods fail. 
A wedgebox analysis can help ascertain whether or not this is the case.
If the wedgebox is a simple box, then a mass reconstruction analysis
can proceed with confidence\footnote{It is true that at the Snowmass Benchmark
point SPS1a \cite{Allanach:2002nj}, a simple box does describe the wedgebox topology 
of a few processes studied \emph{at this point}.  
But these SPS1a studies \cite{Gjelsten:2004ki,Gjelsten:2005aw,Gjelsten:2005sv,Gjelsten:2005vv,Gjelsten:708246,Gjelsten:2006as,Tovey:2008ui,Nojiri:2003tu} are certainly not representative of many other perfectly allowable parameter set choices and/or signature selections.  There seems to have been some tendency to exaggerate, or at least extrapolate in an unsubstantiated way, the proven usefulness of various approaches}.  However, over much of the allowable MSSM parameter space the topology of the wedgebox plot is not merely a lone box --- if a wedge or composite structure is observed, then selecting events from the legs of the wedge or the outer areas generally proves the most effective.  The outermost (more to the right, more toward the top) `incorporated몶 portions of a wedgebox plot basically yield the purest data set.  Here `incorporated portion몶 refers to a clearly delineated zone in the plot in agreement with the predictions of the underlying model.  Thus, in Fig.~\ref{t2-wedge} for instance, straggler events beyond $M \simeq 183\, \hbox{GeV}$, which are not due to neutralino decays, may be eliminated.  Choosing the outermost portion must be weighed against the falling number of events populating such regions (recall though that here MST requirements are relatively modest, and the need for purity generally dominates).  The triangular distribution of events within the dilepton mass distributions imply that an adequately-populated portion of the wedgebox plot will be well-delineated along its outer edges --- which may be taken in this MSSM analysis as indicative that the events in this region are largely from neutralino processes rather than from other process.

  Note that selecting events from the legs of a wedge runs counter to the choices made in all the N-MST and C-MST publications.
In \cite{Nojiri:2007pq} for instance, some care is taken to describe the desirability of so-called
`symmetric events' --- where both legs from the original parent particle contain the same intermediate particle states. The present work, on the other hand, makes the case that the benefits from using un-symmetric decay legs, e.g. efficient isolation of events with the same decay chain structure, may well trump the convenience of symmetric events in the subsequent MST analysis, and therefore unsymmetric events should not be ignored or viewed as an unnecessary complication.
 
The time scale required to collect a sufficient number of events to generate a
wedgebox diagram is roughly comparable to that needed to perform an MST
analysis.  This is in spite of the fact that the wedgebox technique relies upon populating scatter plots while an MST analysis \emph{in principle}
only requires collection of enough events to simultaneously solve the
requisite equations.  \emph{In practice}, ambiguity in assigning the leptons
and multiple solutions to the resulting quartic equation (see bulleted items
in Secn.\ \ref{sec:noj} ) as well as experimental factors (also enumerated
earlier) necessitate a far larger sample of events to perform either of the
MST analyses discussed.  Further, and even more compellingly, without
augmentation by the wedgebox technique\footnote{And/or some other methodology
yielding comparable information.  The wedgebox technique does offer easily
interpretable results, and the di-particle invariant mass distributions the
technique relies upon have events which  cluster near the endpoints, {\it i.e.}, they have triangular distributions, which aid in obtaining clear results.
Here it is perhaps worth noting that other invariant mass combinations or functions proposed which seem in theory to differentiate among event types are effectively of little use if the key region of the distribution is not adequately populated.
  The authors are unaware of other proposed equivalent techniques specifically designed for ascertaining which MSSM production and decay modes are represented in an experimental data set culled by excluded SM event types.}, applying an MST analysis to a quite limited data set is tantamount to wild speculation as to 
what SUSY channels are actually present and the results of such an analysis must be viewed most cautiously.

Scanning over the CM $p_T$ in a $\pm 10\, \hbox{GeV}$-window can also 
enhance the data analysis.
Assumptions that the partonic CM has no transverse momentum
(as implied by equations (\ref{con2}) and (\ref{eqn:gunpt}) ) are basically
incorrect; while in the N-MST method this does not seem to matter,
the C-MST method is much more sensitive to this parameter.
An `averaging' over $p_T$ improves the result, but perhaps a more detailed
analysis should eventually be performed as the latest set of structure
functions and other knowledge of QCD becomes available.

The MST analyses presented here also assume that the decay chain involved is 
a series of two-body decays via intermediate on-mass-shell sleptons.  This 
need not be the case, and the on-mass-shell assumption should be tested.  
This however is beyond the purview of the wedgebox technique.
The di-lepton distribution shapes for on- and off-shell decays are not 
identical \cite{Bachacou:1999zb,Bisset:2005rn},
and this could be used to distinguish between the two possibilities;
however, the effects of cuts, backgrounds and a finite-sized data set
must be considered.  Ref.\ \cite{Chouridou:688033} notes that distribution shapes
for on-shell (sequential two-body decays) and off-shell (three-body decays)
are effectively indistinguishable for some parameter choices.  Also,
Ref.\ \cite{Kitano:2006gv} finds that the shape of the di-lepton distribution
may be affected by the nature of the neutralinos (the extent to which they are
gauginos or higgsinos), illustrating how dynamical issues arising from the
nature of the coupling involved in a decay may not be separable from purely
kinematic issues associated with the relevant masses.

So an alternative to a straight-forward examination of the di-lepton
distribution shape is desirable \cite{Lester:2006cf}.  The Decay Kinematic 
(DK) technique \cite{Kersting:2009ne,Kang:2009sk,Kang:2010hb} might offer such an alternative wherein cross-correlating different invariant mass
distributions resolves the on-shell {\it vs.} off-shell issue, though 
further studies are warranted. Another idea was put forward in 
\cite{Peskin:2008nw}\footnote{See pages 50--51 therein.}, wherein a rudimentary
sketch of a very Dalitz-esque technique to look for the presence of
two-body decay chains is presented --- a realistic study applying this idea 
would be interesting.  
Refs.\ \cite{Lester:2006cf,Lester:2005je} instead champion a `Markov chain'
approach to analyzing the event sample where ``no assumption is made
about the processes causing the observed endpoints.''  Supposedly then
the issue of whether the sleptons are on-mass-shell or off-mass-shell is
rendered moot to the more modest goal of determining a region of parameter
space consistent with the data in a non-MST analysis\footnote{Though
suggested, this issue is not explored in any detail in either of these works.}.

The present work may be thought of as an early installment of a
much grander programme to fuse all known kinematic mass reconstruction
methods together.
The make-up of this programme consists not only of combining fits from
different methods for a static event sample set, but also of improving the
composition of the event set under consideration.
In the present work where the wedgebox technique is used to select (to the 
degree possible) events due to a \emph{specific}
$\widetilde{\chi}_i^0 \widetilde{\chi}_j^0$ neutralino 
pair\footnote{Situations in which $m_{\widetilde{e}} \ne m_{\widetilde{\mu}}$
may also be amenable to such analyses.}.
Once a fairly homogeneous event sample is obtained\footnote{An alternative
track is attempted in Ref.\ \cite{Lester:2005je}, wherein the idea is to
deal with all of the complexity of a mixed data set in the mass analysis
program, rather than bifurcate the analysis into a purifying stage and then
an analysis stage.  The inherent weakness of this approach is that results
from studying just the simplest subset of the events are impeded by the need
to disentangle more confusing events.} then it becomes straightforward to apply various mass reconstruction methods and cross-check them.
For example, in the case of Higgs boson decay considered here, one could try
matching the N-MST results presented here to those from a study of 4-lepton
invariant mass endpoints \cite{Huang:2008qd} --- it would be especially instructive
to compare results at the SPS1a parameter point, for example, where both
techniques, in the total ignorance of sparticle masses,
give poor results individually, but may give a stronger result in
unison\cite{WIP2}.

An MST analysis is then an attractive option if enough mass-shell conditions
can be found to match the number of mass components of the invisible
final-state particles, as is the case for the LSP-generating SUSY decay chains
considered herein.
Further though, the present work shows how endpoint information funneled
through the wedgebox technique can positively supplement such an MST
analysis, as in the augmented C-MST study presented in Secn.\  \ref{sec:cheng}.
No mass-reconstruction technique is immune from possessing potentially
faulty assumptions, and so coupling several complementary analysis techniques
will in general improve reliability as well as accuracy\footnote{Also
minimizing overlapping information content between the analysis components
will increase efficiency.  Whether or not this is a significant issue
would depend on how cpu-intensive the techniques are and on the computing
resources available.}.

Likewise, consideration of suitable inclusive variables, such as
the $m_{T2}$ variable\cite{Lester:1999tx,Barr:2003rg,Lester:2007fq} and its variants \cite{Barr:2002ex,Serna:2008zk}, to augment
either an MST study \cite{Nojiri:2007pq} or an endpoint analysis
\cite{Allanach:2000kt,Tovey:2003ef,Borjanovic:2005tv,Tovey:2008ui,Nojiri:2008hy,Cho:2007dh,Cho:2007qv,Gripaios:2007is,Barr:2007hy,Ross:2007rm} has been shown to
be beneficial, at least in some cases.  Then there is the entire array of
\emph{dynamical} (and thus model-dependent) information associated with
cross-sections and the shapes(spread) of events plotted against
one(or more) parameters.  As noted earlier, kinematics can never really be
totally divorced from the present dynamics.  MSSM/mSUGRA studies combining
information from cross-sections \cite{Lester:2005je} or distribution shapes
\cite{Ross:2007rm,Gjelsten:2006tg,Lester:2006yw} with that from an endpoint analysis have also been
performed and are no doubt the vanguard of many more such studies,
at least if initial LHC results prove favorable.
And, when those first major blocks of data from the LHC become available,
application of numerous analysis techniques ---  including the wedgebox
techniques, would be a good idea.

\section{Acknowledgements}
This work was supported in part by the National Natural Science Foundation of China
Grant No. 10875063 to MB and RL.

\bibliography{wedge}

\end{document}